\documentclass[11pt,a4paper]{article}
\pdfoutput=1
\usepackage{jheppub}
\usepackage{cite}

\usepackage{multirow, graphicx,amssymb,url,mathrsfs,amsmath}
\usepackage{wrapfig,boxedminipage,subfigure,epsfig}
\usepackage{amsxtra,amstext,latexsym,dsfont,amsfonts}
\usepackage{color}
\usepackage[dvipsnames]{xcolor}
\usepackage{float}
\usepackage{slashed}
\usepackage{calligra}
\DeclareFontShape{T1}{calligra}{m}{n}{<->s*[2.2]callig15}{}
\DeclareMathAlphabet{\mathcalligra}{T1}{calligra}{m}{n}

\newcommand{\be}{\begin{equation}}
\newcommand{\ee}{\end{equation}}
\newcommand{\bea}{\begin{eqnarray}}
\newcommand{\eea}{\end{eqnarray}}

\newcommand{\mt}[1]{\textrm{\tiny #1}}

\newcommand{\dd}{\mathrm{d}}

\renewcommand{\Re}{{ \rm Re}}
\renewcommand{\Im}{{ \rm Im}}

\newcommand{\nr}{{\mathcalligra{r}}}

\title{Pole-skipping of scalar and vector fields in hyperbolic space:  conformal blocks and holography}

\author[]{Yongjun Ahn,}
\author[]{Viktor Jahnke,}
\author[]{Hyun-Sik Jeong,}
\author[]{Keun-Young Kim,}
\author[]{Kyung-Sun Lee,}
\author[]{and Mitsuhiro Nishida}

\affiliation[]{School of Physics and Chemistry, Gwangju Institute of Science and Technology, 123 Cheomdan-gwagiro, Gwangju 61005, Korea}

\emailAdd{yongjunahn619@gmail.com}
\emailAdd{viktorjahnke@gist.ac.kr}
\emailAdd{hyunsik@gm.gist.ac.kr}
\emailAdd{fortoe@gist.ac.kr}
\emailAdd{kyungsun.cogito.lee@gmail.com }
\emailAdd{mnishida@gist.ac.kr}

\abstract{Motivated by the recent connection between pole-skipping phenomena of two point functions and four point out-of-time-order correlators (OTOCs), we study the pole structure of thermal two-point functions in $d$-dimensional conformal field theories (CFTs) in hyperbolic space. We derive the pole-skipping points of two-point functions of scalar and vector fields by three methods (one field theoretic and two holographic methods) and confirm that they agree. We show that the leading pole-skipping point of two point functions is related with the late time behavior of conformal blocks and shadow conformal blocks in four-point OTOCs.
 }

\begin{document}
\maketitle



\section{Introduction}

Correlation functions are basic objects in quantum field theories (QFTs) since they contain information about physical objects like, for example, scattering cross-sections. A standard definition of correlation functions includes the time-ordering operator, but we can also consider out-of-time-order correlation functions (OTOCs) \cite{larkin1969quasiclassical}. Recently, OTOCs have been established as a fundamental quantity for measuring the delocalization (or scrambling) of quantum information in chaotic quantum many-body systems \cite{Kitaev-2014, Maldacena:2015waa}. In some chaotic systems, a four point OTOC  $\langle V(t, \mathbf{d})W(0,0)V(t, \mathbf{d})W(0,0)\rangle$ behaves as\footnote{Another propagation behavior with  diffusion was observed in \cite{Swingle:2016jdj, Aleiner:2016eni}. See also \cite{Xu:2018xfz,Khemani:2018sdn} for a discussion of other possible OTOC growth forms.} 
\begin{align}
\frac{\langle V(t, \mathbf{d})W(0,0)V(t, \mathbf{d})W(0,0)\rangle}{\langle V(t, \mathbf{d}) V(t, \mathbf{d})\rangle\langle W(0,0)W(0,0)\rangle}\sim1-\varepsilon e^{\lambda_L(t-t_*-\mathbf{d}/v_B)}+\cdots,\label{OTOC}
\end{align}
where $\lambda_L>0$ is the Lyapunov exponent, $\mathbf{d}$ is a large spatial distance between the operators $V$ and $W$, $v_B$ is the butterfly velocity, and $\varepsilon$ is a prefactor which depends on $V, W$\footnote{In holographic CFTs, $\varepsilon$ depends on conformal dimension of $V$ and $W$, i.e., $\varepsilon \propto \Delta_V \Delta_W$ \cite{Roberts:2014ifa, Shenker:2014cwa}.}, and regulators. The exponential growth in (\ref{OTOC}) is a late time behavior before the scrambling time $t_*$. Examples of chaotic systems are QFTs which have a gravitational description, e.g. \cite{Shenker:2013pqa, Shenker:2013yza, Roberts:2014isa, Roberts:2014ifa, Shenker:2014cwa, Polchinski:2016xgd, Perlmutter:2016pkf,  Maldacena:2016hyu, Jensen:2016pah, Maldacena:2016upp, Davison:2016ngz,Jahnke:2018off,Jahnke:2019gxr, Fischler:2018kwt,Avila:2018sqf, Jahnke:2017iwi,Aalsma:2020aib, Geng:2020kxh}.

Interestingly, in holographic theories, it has been proposed that $\lambda_L$ and $v_B$ can be determined also by the retarded {\it two} point function $G^R(\omega, k)$ of energy density $T_{00}$ in momentum space \cite{Grozdanov:2017ajz, Blake:2017ris}. Generally, $G^R(\omega, k)$ can be expressed as $G^R(\omega, k)=\frac{b(\omega, k)}{a(\omega, k)}$, and one can compute $a(\omega, k)$ and $b(\omega, k)$ by using classical solutions of the holographic models. In a large class of holographic models, it was shown that there are points $\omega_*=2\pi iT, k_*=\pm2\pi iT/v_B$ such that $a(\omega_*, k_*)=b(\omega_*, k_*)=0$, where $T$ is the Hawking temperature. These points are related to the  Lyapunov exponent $\lambda_L=2\pi T$ and the butterfly velocity $v_B$ appearing in (\ref{OTOC}). This phenomenon is called ``pole-skipping" because the divergence of $G^R(\omega_*, k_*)$ from $a(\omega_*, k_*)=0$ at the poles is skipped by $b(\omega_*, k_*)=0$.
It has been also proposed that the pole-skipping points correspond to the special points of momentum modes in the Einstein's equations near the black hole horizon \cite{Blake:2018leo}. At these special points, {\it two} independent incoming solutions at horizon are available so the Green's function is not uniquely defined. For a more rigorous as well as intuitive explanation, we refer to section \ref{intro}.

Having this interesting connection between four-point OTOC and pole-skipping phenomena in two-point function of energy density $T_{00}$, it is natural to ask if there is a similar connection  and interesting relevant physics for fields other than $T_{00}$. In particular,  the pole-skipping points for scalar and Maxwell fields in {\it flat} space were computed in \cite{Grozdanov:2019uhi, Blake:2019otz, Natsuume:2019xcy, Natsuume:2019vcv, Wu:2019esr, Abbasi:2019rhy} by an exact calculation of $G^R(\omega, k)$ or by a near-horizon analysis. 
 However, in contrast to the pole-skipping of energy density, the relation between the behavior of four point functions and the pole-skipping points for the two point function of bulk scalar and Maxwell fields is not well-studied.\footnote{In these cases, there is no pole-skipping point in the upper half of the complex $\omega$-plane.}  
 In this work, we try to fill this gap by studying the pole-skipping structure of scalar and vector fields in {\it hyperbolic} space and investigating its relation to conformal block. The hyperbolic space is advantageous because we can compute the Green's function analytically. 
 This study also allows us to uncover some features of how the pole-skipping phenomenon manifests itself when the field theory lives in a curved space.
 For some other recent studies regarding pole-skipping, see \cite{Grozdanov:2018kkt,Li:2019bgc,Natsuume:2019sfp, Ahn:2019rnq, Ceplak:2019ymw, Liu:2020yaf}.

In addition to the holographic computations, the pole-skipping in conformal field theories (CFTs) was also studied in \cite{ Liu:2020yaf, Haehl:2018izb, Das:2019tga, Haehl:2019eae}. The authors of \cite{Haehl:2019eae} calculated the pole-skipping points in conformal two point functions of energy density $T_{00}$ on $S^1\times\mathbb{H}^{d-1}$, where the size of $S^1$ is $\beta=2\pi$, and showed that the relevant pole-skipping point is related to the  Lyapunov exponent $\lambda_L=2\pi/\beta$ and the butterfly velocity $v_B=1/(d-1)$ in the holographic CFTs on hyperbolic space \cite{Perlmutter:2016pkf}.  Since the conformal two point function is universal in any CFTs including free CFTs, the existence of the relevant pole-skipping point in conformal two point functions does not directly indicate the chaotic behavior in CFTs.

A characteristic feature of the holographic CFTs, which is supported by their Einstein gravity dual, is that 
 the sub-leading term in the OTOC (\ref{OTOC}) is effectively approximated by the conformal block with exchange of the energy momentum tensor $T_{\mu\nu}$ \cite{Perlmutter:2016pkf}\footnote{If the theory contains light higher-spin fields, the exchange of higher-spin fields is effectively dominant. In this case, the pole-skipping points of higher-spin fields would be related to the behavior of OTOCs \cite{Haehl:2018izb}. Note, however, that theories with a finite number of light higher-spin fields are pathological \cite{Camanho:2014apa} and break a large gap in the higher-spin single-trace sector for holographic CFTs \cite{Heemskerk:2009pn}.}. In the bulk language, this approximation is understood by the ``graviton dominance" of bulk amplitudes \cite{Cornalba:2006xm}.   Assuming this property,   one can derive $\lambda_L$ and $v_B$ in the holographic CFTs  through  the conformal block with exchange of $T_{\mu\nu}$.  Motivated by this fact, we expect a relation between the conformal block's behavior and pole-skipping in conformal two point functions. It is interesting to check our expectation with general fields other than $T_{\mu\nu}$, especially because the pole-skipping points of bulk scalar and vector fields in flat space were observed by holographic computations. In this paper, we find it also in hyperbolic space.
 
 In this paper, we study a relation between the conformal blocks and the pole-skipping points of two point functions in hyperbolic space with an analytic continuation. Motivated by the Lyapunov exponent $\lambda_L$ and the butterfly velocity $v_B$, we define two exponents in the late time behavior of conformal blocks. These exponents depend on conformal dimension and spin of the exchange operators in conformal blocks as known in the Regge limit. With the exchange of scalar or vector fields, we show that the exponents of conformal blocks and their shadow conformal blocks are related to the leading pole-skipping points of two point functions of the exchange operators in momentum space. In computations of the pole-skipping points, we use three methods: (1) pole-skipping analysis of conformal two point functions in CFTs, (2) pole-skipping analysis of retarded two point functions computed holographically, and (3) near-horizon analysis of bulk classical equations. The pole-skipping points obtained by these three computation methods are consistent with each other and with the exponents in conformal blocks.
 
One of the novel aspects of our study is that the Rindler-AdS$_{d+1}$ geometry allows us to perform a very complete analysis, in which we can explicitly compute the Green's functions in both sides of the AdS/CFT duality, and extract the corresponding pole-skipping points. Moreover, we can check that the same pole-skipping points can be obtained by a simple near-horizon analysis. As far as we know, the only system that is simple enough to allow for such a thorough analysis is the BTZ black hole \cite{Grozdanov:2019uhi, Blake:2019otz}. We show that it holds for an arbitrary number of dimensions in hyperbolic space.

 This paper is organized as follows. In Section \ref{SCB}, we review the analytic continuation of conformal blocks for OTOCs in hyperbolic space and define the two exponents. The pole-skipping points of two point functions are investigated by conformal two point functions in Section \ref{SCTF}, by holographic retarded Green's function in Section \ref{SHGF}, and by analysis of bulk equations of motion near the horizon of the black hole geometry in Section \ref{SNH}. In Section \ref{discussion}, we discuss our results and future work.

\section{Late time behavior of OTOCs  in hyperbolic space from conformal block \label{SCB}}

In this section, we review the OTOC behavior in hyperbolic space at late time by an analytic continuation of the relevant Euclidean conformal blocks. We focus on the cases where the exchange operators are scalar or vector fields, and define ``exponents'' in the late time behavior. In the case where the energy momentum tensor is the dominant exchange operator, these ``exponents'' correspond to the Lyapunov exponent and the butterfly velocity in the holographic CFTs. Our computation is based on \cite{Haehl:2019eae}. See also \cite{Perlmutter:2016pkf, Cornalba:2006xm}.

First, we start reviewing the relation between hyperbolic, Rindler, and Euclidean spaces based on \cite{Haehl:2019eae}. Let us consider a hyperbolic space metric with a periodic Euclidean time $\tau$:
\begin{align}
\label{cftco}
d s^2_{S^1\times \mathbb{H}^{d-1}}=d \tau^2+\frac{1}{\rho^2}\left( d \rho^2+d x^i_\perp d x_{\perp i}\right)\,,
\end{align}
where $d \ge 3$.
This metric is conformally equivalent to a (Euclidean) Rindler space metric
\begin{align}
d s^2_{S^1\times \mathbb{H}^{d-1}}=\frac{1}{\rho^2}\left(\rho^2d \tau^2+ d \rho^2+d x^i_\perp d x_{\perp i}\right)=\frac{1}{\rho^2}d s^2_\textrm{Rindler} \,,
\end{align}
where the period of $\tau$ is set as $\beta = 2\pi$.\footnote{{ Note that we set the length scale, which we call $\ell_\mathrm{AdS}$, to unity i.e. $\ell_\mathrm{AdS} =1$. The subscript AdS comes from the relation with the holographic computations in sections \ref{SHGF} and \ref{SNH}}. When we recover $\ell_\mathrm{AdS}$, we may  recover it as $1 \to 1/\ell_\mathrm{AdS} = 2\pi/\beta = 2\pi T$, where $T$ is the temperature. \label{rec1}}
The Rindler space can be embedded in the Euclidean space $d s^2=\delta_{\mu\nu}d x^\mu d x^\nu$ via 
\begin{align}
x^\mu=\left(\rho\sin \tau, \rho\cos \tau, x^i_\perp\right).\label{ct1}
\end{align}
By this coordinate transformation and a conformal transformation, thermal CFT's correlation functions in the hyperbolic space can be computed by Euclidean conformal blocks \cite{Casini:2011kv}. 

The Euclidean conformal block is a function of the two cross ratios $u$ and $v$,
\begin{align}
    u:=\frac{x^2_{12}x^2_{34}}{x^2_{13}x^2_{24}} \,, \qquad v:=\frac{x^2_{14}x^2_{23}}{x^2_{13}x^2_{24}} \,,
\end{align}
which are defined by the distances $x^2_{ab}:=\delta _{\mu\nu}(x_a-x_b)^\mu(x_a-x_b)^\nu$ in the Euclidean space. The distances  $x^2_{ab}$ are related to the Rindler coordinate as
\begin{align}
x^2_{ab}=&2\rho_a\rho_b[\cosh \mathbf{d}(a, b)-\cos(\tau_a-\tau_b)]\,,\\
\cosh \mathbf{d}(a, b):=&\frac{\rho^2_a+\rho^2_b+(x_{\perp a}-x_{\perp b})^2}{2\rho_a\rho_b}\,.\label{dd}
\end{align}

We want to consider a four-point OTOC of pairwise equal scalar  operators\footnote{We consider the scalar case for convenience and simplicity. For other cases such as vector or tensor operators, the expressions of conformal blocks will be more complicated.} $V$ and $W$ at coincident points, $\langle V(t, \mathbf{d})W(0,0)V(t, \mathbf{d})W(0,0)\rangle$.  If we denote the arguments of $V$ by 1 and 2 while the ones of $W$ by 3 and 4, then $\mathbf{d}(1,2)=\mathbf{d}(3,4)=0$ and $\mathbf{d}(1,3)=\mathbf{d}(2,3)=\mathbf{d}(1,4)=\mathbf{d}(2,4)=:\mathbf{d}>0$. The real time $t$ is obtained by an analytic continuation of the Euclidean time:  $\tau_a=it_a+\delta_a$ with $t_1=t_2=t$ and $t_3=t_4=0$. Here, for the OTOC configuration, we take $\delta_a \to 0$ with $\delta_1> \delta_3 >\delta_2 >\delta_4$. In this configuration, we have the cross ratios $u$ and $v$~\cite{Haehl:2019eae}
\begin{align}
u=&\frac{x^2_{12}x^2_{34}}{x^2_{13}x^2_{24}}=\frac{[1-\cos \delta_{12}][1-\cos \delta_{34}]}{[\cosh \mathbf{d}-\cosh (t-i\delta_{13})][\cosh \mathbf{d}-\cosh (t-i\delta_{24})]} \,,\label{acu}\\
v=&\frac{x^2_{14}x^2_{23}}{x^2_{13}x^2_{24}}=\frac{[\cosh \mathbf{d}-\cosh (t-i\delta_{14})][\cosh \mathbf{d}-\cosh (t-i\delta_{23})]}{[\cosh \mathbf{d}-\cosh (t-i\delta_{13})][\cosh \mathbf{d}-\cosh (t-i\delta_{24})]} \,,\label{acv}
\end{align}
where $\delta_{ab}:=\delta_a-\delta_b$.

Let us now consider the relation between the four point function and the conformal blocks $G^{(\ell)}_{\Delta}(u, v)$ associated with the exchange of different primary
operators $\mathcal{O}$ and their descendants~\cite{Haehl:2019eae}:
\begin{align}
  \langle V(x_1)V(x_2)W(x_3)W(x_4)\rangle=\frac{1}{x_{12}^{2\Delta_V}x_{34}^{2\Delta_W}}\sum_\mathcal{O}C_{VV\mathcal{O}}C_{WW\mathcal{O}}G^{(\ell)}_{\Delta}(u, v)\,,
\end{align}
where $\ell$ and $\Delta$ are spin and conformal dimension of the exchange operators, which are scalars, vectors, or symmetric and traceless tensors\footnote{Other operators such as fermion and anti-symmetric tensor are forbidden because of symmetry. {When $\ell$ is not an even integer, the OPE coefficient for three point function $\langle V(x_1)V(x_2)\mathcal{O}_{\nu_1\cdots\nu_\ell}(x)\rangle$ vanishes \cite{Simmons-Duffin:2016gjk}, where $\mathcal{O}_{\nu_1\cdots\nu_\ell}(x)$ is a primary operator with spin $\ell$. Even in this case, the conformal block can be constructed because it does not depend on the OPE coefficient.}}. For convenience, let us introduce a notation 
\begin{equation}
     \langle V(x_1)V(x_2)W(x_3)W(x_4)\rangle_{\ell, \Delta} \sim G^{(\ell)}_{\Delta}(u,v) \,,
\end{equation}
which represents the contribution from a given exchange operator with spin $\ell$ and conformal dimension $\Delta$.

\begin{figure}
\centering
     {\includegraphics[width=8cm]{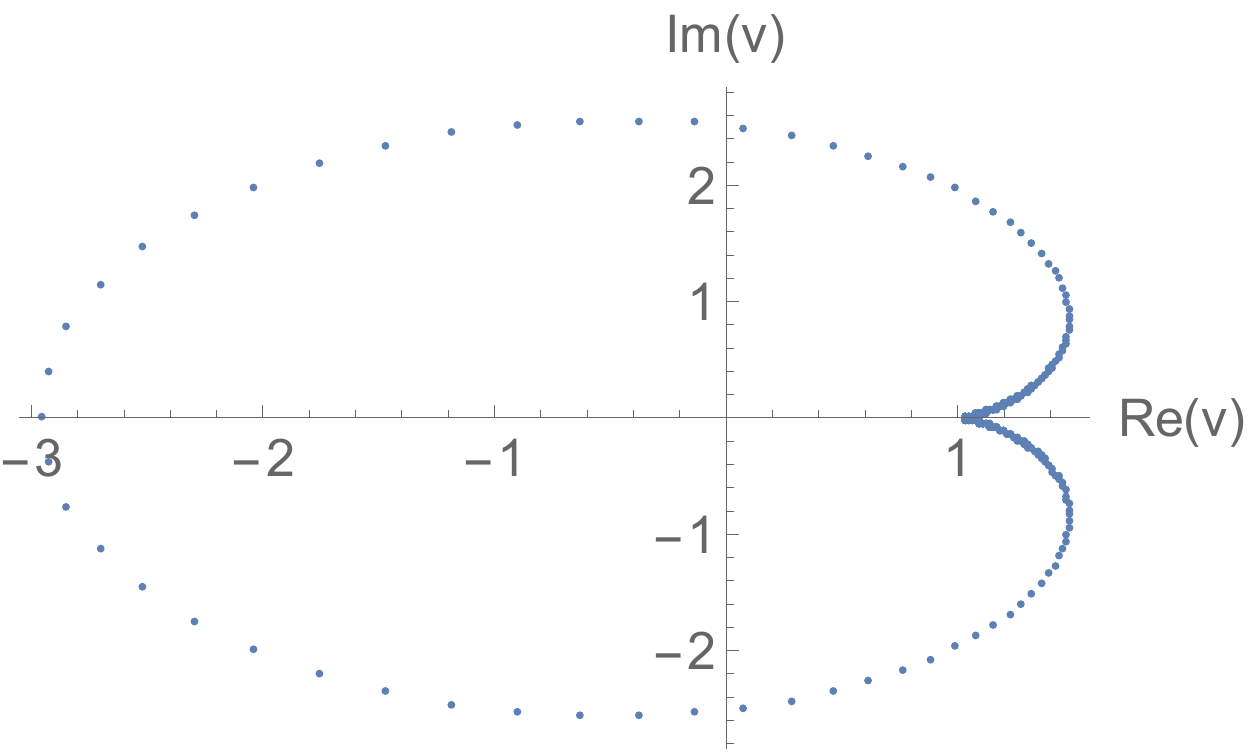}}
 \caption{Plot of $v$ with $\mathbf{d}=10,\ \delta_1=0.6, \ \delta_3=0.4, \ \delta_2=0.2, \ \delta_4=0$. The horizontal axis is $\Re[v]$, and the vertical axis is $\Im[v]$. We plot $v$ from $t=8$ to $t=12$.}\label{v}
\end{figure}

In particular, the OTOC, $\langle V(t, \mathbf{d})W(0,0)V(t, \mathbf{d})W(0,0)\rangle$ 
 can be obtained by an analytic continuation ($v\to e^{-2\pi i}v$) of the Euclidean conformal block~\cite{Roberts:2014ifa}:
\begin{equation} \label{jhkn}
    \langle V(t, \mathbf{d})W(0,0)V(t, \mathbf{d})W(0,0)\rangle_{\ell,\Delta}
    \sim G^{(\ell)}_{\Delta}(u, v \to e^{-2\pi i}v) \,.
\end{equation}
The analytic continuation $v\to e^{-2\pi i}v$ corresponds to the time evolution of $v$.  Under the time evolution from $t=0$, $v$ rotates around $v=0$ clock-wisely as shown in Figure \ref{v}, where we considered the OTOC configuration $\delta_1> \delta_3 >\delta_2 >\delta_4$. At two limits $t\gg\mathbf{d}$ or $\mathbf{d}\gg t$, $v$ goes to $1$. The time evolution of $v$ around $t\sim \mathbf{d}$ is nontrivial. To show it clearly we chose the time range $8< t <12$.

We want to study the behavior of \eqref{jhkn} in the late time and large distance limit, $t\gg \mathbf {d}\gg 1$, since our goal is to identify two exponents similar to the ``Lyapnov exponent'' and ``butterfly velocity''. This limit corresponds to 
the following limit of $u$ and $v$: 
\begin{equation} \label{ghjnc}
u\sim e^{-2t}\delta^2_{12}\delta^2_{34} \sim 0, \qquad
v\sim 1+e^{-t+\mathbf{d}}\delta_{12}\delta_{34} \sim 1,
\end{equation}
which is obtained from (\ref{acu}) and (\ref{acv}) by assuming  $\delta_{ab}\ll1$ for the OTOC configuration and $t\gg \mathbf {d}\gg 1$.

First, if $u \sim 0$  the Euclidean conformal block can be expressed in terms of the formula \cite{Dolan:2011dv}
\begin{align}
G^{(\ell)}_{\Delta}(u, v)\sim u^{\frac{1}{2}(\Delta-\ell)}(1-v)^{\ell}\;_{2}F_1\left(\frac{1}{2}(\Delta+\ell), \frac{1}{2}(\Delta+\ell); \Delta+\ell; 1-v\right) \;\;\;\;\;\;\;  (u\to0)\,.
\end{align}
Note that, from \eqref{acu}, the $u \to 0$ limit can be obtained in a large spatial distance limit,  $\mathbf{d} \gg 1$ at any $t$ without imposing the large $t$ limit.  
 Since the hypergeometric function $_2F_1$ is a multi-valued function of $v$, we can pick up the monodromy along $v\to e^{-2\pi i}v$. The hypergeometric function $_{2}F_1\left(\frac{1}{2}(\Delta+\ell), \frac{1}{2}(\Delta+\ell);  
\Delta+\ell;  1-v\right)$ is a fundamental solution of the hypergeometric differential equation, and another fundamental solution is given by\footnote{Strictly speaking, they are fundamental solutions when $\Delta+\ell$ is not an integer. However, even when $\Delta+\ell\in \mathbb{Z}$, there is a fundamental solution whose leading order is $(1-v)^{1-\Delta-\ell}$. See, for example, \cite{NIST:DLMF}.} $(1-v)^{1-\Delta-\ell}\;_{2}F_1\left(1-\frac{1}{2}(\Delta+\ell), \right. $ $\left. 1-\frac{1}{2}(\Delta+\ell); 2-(\Delta+\ell); 1-v\right)$. Thus, the monodromy can be expressed as
\begin{align}
&_{2}F_1\left(\frac{1}{2}(\Delta+\ell), \frac{1}{2}(\Delta+\ell); \Delta+\ell; 1-e^{-2\pi i}v\right)-\;_{2}F_1\left(\frac{1}{2}(\Delta+\ell), \frac{1}{2}(\Delta+\ell); \Delta+\ell; 1-v\right)\notag\\
=&A\;_{2}F_1\left(\frac{1}{2}(\Delta+\ell), \frac{1}{2}(\Delta+\ell); \Delta+\ell; 1-v\right)\notag\\
+&B(1-v)^{1-\Delta-\ell}\;_{2}F_1\left(1-\frac{1}{2}(\Delta+\ell), 1-\frac{1}{2}(\Delta+\ell); 2-(\Delta+\ell); 1-v\right),\label{hgf3}
\end{align}
where $A$ and $B$ are constants. The explicit forms of $A$ and $B$ when $\Delta+\ell$ is not an integer are derived in Appendix \ref{mhgf}.

Next, if $v \sim 1$, for $\Delta +\ell>1$, the second term in  (\ref{hgf3}) is dominant, and it behaves as $(1-v)^{1-\Delta-\ell}$. By using the dominant behavior in the monodromy (\ref{hgf3}), we obtain the analytic continuation of the conformal block in the OTOC
\begin{align}
&G^{(\ell)}_{\Delta}(u, v \to e^{-2\pi i}v)|_{u\sim 0, v\sim 1}
\sim u^{\frac{1}{2}(\Delta-\ell)}(1-v)^{1-\Delta}
\sim e^{(\ell-1)t-(\Delta-1)\mathbf{d}},\label{mcb}
\end{align}
where we used \eqref{ghjnc}. 
One can also derive the late time behavior (\ref{mcb}) from a formula of the conformal block in the Regge limit \cite{Perlmutter:2016pkf, Cornalba:2006xm}.

We define two exponents $\omega_*$ and $L_*^+$ by using (\ref{mcb}) as
\begin{align}
\langle V(t, \mathbf{d})W(0,0)V(t, \mathbf{d})W(0,0)\rangle_{\ell,\Delta} \sim e^{(\ell-1)t-(\Delta-1)\mathbf{d}}=: e^{-i\omega_* t+L_*^+\mathbf{d}}.\label{dol}
\end{align}
 In particular, $\omega_*$ and $L^+_*$ for the exchange operator with spin $\ell$ and conformal dimension $\Delta$ are
 \begin{align}
 \omega_*=i(\ell-1), \qquad L^+_*=1-\Delta.\label{eol}
 \end{align}
For the shadow operator (see, for example, \cite{Ferrara:1972uq, SimmonsDuffin:2012uy}) with spin $\ell$ and conformal dimension $d-\Delta$, $\omega_*$ and $L^-_*$ {are defined as
 \begin{align}
 \omega_*=i(\ell-1), \qquad L^-_*=\Delta-d+1\,,\label{eols}
 \end{align}
 where we use the replacement $\Delta\to d-\Delta$ in the exponents  (\ref{eol})\footnote{By replacing $\Delta \to d - \Delta$ we have a condition $d-\Delta+\ell>1$ for the second term of \eqref{hgf3} to be dominant. Thus, together with $\Delta + \ell > 1$, our results are valid for $d+\ell-1>\Delta>1-\ell$ if the OTOC is related to the dominant term of \eqref{hgf3}.}. 
Note that $\omega_* = 2\pi T i (\ell -1)$ if we recover $\ell_\mathrm{AdS}$ as explained in footnote \ref{rec1}. 
In the next section, we will show that the exponents (\ref{eol}) and (\ref{eols}) with $\ell=0, 1$ are related to the relevant pole skipping points of the corresponding two point functions. 

Before computing the pole-skipping points, let us discuss why the shadow conformal block is also important in the pole-skipping by using the shadow formalism. Define the shadow operator $\widetilde{\mathcal{O}}^{\mu_1\cdots\mu_\ell}(x)$ of $\mathcal{O}^{\nu_1\cdots\nu_\ell}(x)$ and the projection operator $|\mathcal{O}|$ \cite{ Haehl:2019eae, Dolan:2011dv, SimmonsDuffin:2012uy} 
\begin{align}
\widetilde{\mathcal{O}}^{\mu_1\cdots\mu_\ell}(x):&=\frac{k_{\Delta, \ell}}{\pi^{d/2}}\int d^d y\frac{\prod_{i=1}^\ell\left(\delta^{\mu_i\nu_i}(x-y)^2-2(x-y)^{\mu_i}(x-y)^{\nu_i}\right)}{(x-y)^{2(d-\Delta+\ell)}}\mathcal{O}_{\nu_1\cdots\nu_\ell}(y),\\
|\mathcal{O}|:&=\frac{k_{d-\Delta, \ell}}{C_\mathcal{O}\pi^{d/2}}\int d^d \xi \;\mathcal{O}^{\mu_1\cdots\mu_\ell}(\xi)|0\rangle\langle0| \widetilde{\mathcal{O}}_{\mu_1\cdots\mu_\ell}(\xi)\notag\\
&=\frac{k_{d-\Delta, \ell}}{C_\mathcal{O}\pi^{d/2}}\int d^d \xi \;\widetilde{\mathcal{O}}^{\mu_1\cdots\mu_\ell}(\xi)|0\rangle\langle0| \mathcal{O}_{\mu_1\cdots\mu_\ell}(\xi),
\end{align}
where $k_{\Delta, \ell}:=\frac{\Gamma(\Delta-1)\Gamma(d-\Delta+\ell)}{\Gamma(\Delta+\ell-1)\Gamma(\Delta-d/2)}$, and $C_\mathcal{O}$ is a normalization constant of $\mathcal{O}$'s two point function. Note that the projection operator satisfies $|\mathcal{O}|^2=|\mathcal{O}|$. By inserting the projection operator into the conformal four point function, one can obtain
\begin{equation}\label{IFP}
\begin{split}
&\langle V(x_1)V(x_2)|O|^2W(x_3)W(x_4)\rangle\\
=&\left(\frac{k_{d-\Delta, \ell}}{C_\mathcal{O}\pi^{d/2}}\right)^2\int \int d^d \xi d^d \xi'   \ \langle V(x_1)V(x_2) \widetilde{\mathcal{O}}^{\mu_1\cdots\mu_\ell}(\xi) \rangle  \ \langle \mathcal{O}_{\mu_1\cdots\mu_\ell}(\xi) \mathcal{O}_{\nu_1\cdots\nu_\ell}(\xi')\rangle  \\
& \qquad \qquad\qquad\qquad \qquad\quad \ \times
\langle \widetilde{\mathcal{O}}^{\nu_1\cdots\nu_\ell}(\xi')W(x_3)W(x_4)\rangle \\
=& \ \langle V(x_1)V(x_2)\rangle\langle W(x_3)W(x_4)\rangle\left(A_1G^{(\ell)}_{\Delta}(u, v)+A_2G^{(\ell)}_{d-\Delta}(u, v)\right) \,,
\end{split}
\end{equation}
where $A_i$ are constants.
Note that (\ref{IFP}) is a solution of the conformal Casimir equation due to the three point functions in the second term.  Thus, it can be written as a linear combination of conformal block and its shadow conformal block, which is the term in the last line. 
 Therefore, we expect that there is a contribution of the pole-skipping  structure of $\mathcal{O}$'s two point function to the shadow conformal block. In fact, for projecting out the unphysical shadow conformal block, the authors of \cite{Haehl:2019eae} conjectured that one of the leading pole-skipping points in the energy momentum tensor two point function corresponds to the ``physical" pole in the two point function of the shadow tensor  mode.

\section{Pole-skipping analysis: conformal  two point functions\label{SCTF} in hyperbolic space}\label{section3}

In this section, we study the pole-skipping points of conformal two point functions of scalar and vector fields in $S^1\times \mathbb{H}^{d-1}$. 
Following \cite{Haehl:2019eae,Ohya:2016gto}, they can be computed by the embedding space formalism i.e. by embedding  $S^1\times \mathbb{H}^{d-1}$ in $\mathbb{R}^{1,{d+1}}$.
From the explicit expression of Fourier transformed two point functions of scalar and vector fields, we can investigate the pole-skipping structure. In the last subsection, we check that the leading pole-skipping points are related to the exponents (\ref{eol}) and (\ref{eols}).

\subsection{Scalar field}

We review the computation of scalar two point function in momentum space \cite{Haehl:2019eae, Ohya:2016gto}. To start with, we briefly introduce the embedding formalism (see, for example, \cite{Costa:2011mg}) which embeds the $d$-dimensional Euclidean space in the lightcone of $d+2$-dimensional Minkowski spacetime. Its basic idea is that the conformal group of $d$-dimensional Euclidean space $SO(d+1,1)$ can be linearized as the isometry group of $d+2$-dimensional Minkowski spacetime. The embedding formalism also makes the computations of two point functions of fields with spin $\ell=1,2,\dots$ more accessible as we will see in Subsection \ref{tau2pt}. From now on, we follow the conventions of \cite{Haehl:2019eae}. 

Using Rindler coordinates (\ref{ct1}), we can embed the coordinates in $S^1\times \mathbb{H}^{d-1}$ as
\begin{equation}
    P^A=\left(\frac{1+x^2}{2\rho},\frac{1-x^2}{2\rho},\frac{x^\mu}{\rho}\right)=\left(\frac{1+\rho^2+x_\perp^2}{2\rho},\frac{1-\rho^2-x_\perp^2}{2\rho},\sin\tau,\cos\tau,\frac{x_\perp^i}{\rho}\right),
\end{equation}
which lie on the lightcone $P^2=0$, and the indices are $A=(I,II,\mu),\; \mu=(0,1,i)$ with the transverse index $i$ ranging from $2$ to $d-1$. Then the scalar two point functions in terms of the embedding coordinates are
\begin{equation} \label{green098}
\begin{split}
\mathcal{G}^\Delta(P_1,P_2)&\equiv\langle\phi^\Delta(P_1)\phi^\Delta(P_2)\rangle_{S^1\times\mathbb{H}^{d-1}} \\ 
&=\frac{1}{(-2P_1\cdot P_2)^\Delta}=\frac{1}{(-2\cos(\tau_{1}-\tau_2)-2Y_1\cdot Y_2)^\Delta}\,,
\end{split}
\end{equation}
where $(-2P_1\cdot P_2)=(x_1-x_2)^2$, $\phi^\Delta$ is a primary scalar field with conformal dimension $\Delta$, and 
\begin{equation}
Y^A:=\left(\frac{1+\rho^2+x_\perp^2}{2\rho},\frac{1-\rho^2-x_\perp^2}{2\rho},0,0,\frac{x_\perp^i}{\rho}\right) \,,
\end{equation}
in terms of which the geodesic distance (\ref{dd}) is expressed as
\begin{equation}
 \cosh \mathbf{d}(1, 2)=-Y_1\cdot Y_2   \,.
\end{equation}

To perform the Fourier transformation of the parameterized scalar two point function on $S^1\times\mathbb{H}^{d-1}$, we need the eigenfunctions of the scalar Laplacian $\square_{S^1\times\mathbb{H}^{d-1}}$: 
\begin{align}
\square_{S^1\times\mathbb{H}^{d-1}}=\partial_\tau^2+\rho^2\partial_\rho^2-(d-3)\rho\partial_\rho +\rho^2 \square_{\mathbb{R}^{d-2}}.\label{eesc}
\end{align}
The eigenfunction is
\begin{equation} 
    f(P;\omega_E, k,\vec p_\perp)\propto\rho^{\frac{d-2}{2}}K_{ik}(|p_\perp|\rho)e^{i(\omega_E\tau+\vec p_\perp\cdot \vec x_\perp)}\label{eigen1},
\end{equation}
with the eigenvalue
\begin{equation} \label{eigen123}
    -\omega_E^2-k^2-\left(\frac{d-2}{2}\right)^2\,.
\end{equation}
Here, $K_{ik}$ is the modified (hyperbolic) Bessel functions of the second kind  and the momentum space conjugate to $(\tau,\rho,x_\perp^i)$ is denoted as $(\omega_E,k,p_\perp^i)$. 
By using the eigenfunction \eqref{eigen1} and the scalar two point function \eqref{green098}, we can obtain the Fourier transformed two point function as 
\begin{equation}
\begin{split}\label{erase}
\mathcal G^\Delta&(\omega_E,k)f(P;\omega_E, k,\vec p_\perp) \\
&=\int dP'\, \mathcal G^{\Delta}(P,P')\,f(P';\omega_E, k,\vec p_\perp)  \\
&=\int dP' \frac{1}{(-2P\cdot P')^{\Delta}}\rho'^{\frac{d-2}{2}}K_{ik}(|\vec p_\perp|\rho')e^{i\omega_E\tau'+i\vec p_\perp\cdot \vec x_\perp'}\\
&=\frac{\pi^{\frac{d}{2}}}{\Gamma(\Delta)}\frac{f(P;\omega_E, k,\vec p_\perp)|\Gamma(\alpha)|^2}{\Gamma(\alpha+\alpha^*+\frac{d}{2}-\Delta)}\lim\limits_{z\rightarrow1^-}\;_2F_1\left(\alpha,\alpha^*;\alpha+\alpha^*+\frac{d}{2}-\Delta;z\right)\,,
\end{split}
\end{equation}
where $\alpha:= \frac{1}{2}(\omega_E+ik-\frac{d-2}{2}+\Delta)$. For the details and some subtleties of the explicit integration, see {appendix \ref{integration}} and \cite{Haehl:2019eae, Ohya:2016gto}. 
Thus, we obtain the scalar two point function:
\begin{equation}
	\mathcal{G}^\Delta(\omega_E,k)=\frac{\pi^\frac{d}{2}}{\Gamma(\Delta)}\frac{|\Gamma(\alpha)|^2}{\Gamma(\alpha+\alpha^* -\delta)}\lim_{z\rightarrow1^-}\;_2F_1\left(\alpha,\alpha^*;\alpha+\alpha^*-\delta\,;z\right)\,, \label{fttf11}
\end{equation}
where we defined 
\begin{equation}
 \delta := \Delta - d/2   \,. 
\end{equation}

As explained in detail in appendix \ref{regularize}, as far as the pole-skipping structure is concerned, the hypergeometric function can be expressed in two ways depending on whether $\delta$ is a non-negative integer or not, i.e. $\delta \in \mathbb{Z}^{*+} := \{ 0 \} \cup \mathbb{Z}^+$ or $\delta  \notin \mathbb{Z}^{*+}$. We find that, if $\delta \notin \mathbb{Z}^{*+}$ \eqref{rep1},
\be
\;_2F_1(\alpha,\alpha^*;\alpha+\alpha^*-\delta;1)\rightarrow\frac{\Gamma(\alpha+\alpha^*-\delta)\Gamma(-\delta)}{\Gamma(\alpha-\delta)\Gamma(\alpha^*-\delta)} \,, \label{hyper2gamma}
\ee
and if $\delta \in \mathbb{Z}^{*+}$
\eqref{rep2}
\begin{equation}
\;_2F_1(\alpha,\alpha^*;\alpha+\alpha^* -\delta;1) \rightarrow\frac{\Gamma(\alpha+\alpha^*-\delta)\Gamma(-\delta)}{\Gamma(\alpha-\delta)\Gamma(\alpha^*-\delta)} \left[\psi(\alpha)+\psi(\alpha^*)\right]  \,.
\end{equation}

Therefore, the scalar two point function finally becomes 
\begin{equation}
\mathcal{G}^\Delta(\omega_E,k)\propto\frac{\Gamma(\frac{1}{2}(\omega_E+ ik+\delta+1))\Gamma(\frac{1}{2}(\omega_E- ik+\delta+1))}{\Gamma(\frac{1}{2}(\omega_E+ ik-\delta+1))\Gamma(\frac{1}{2}(\omega_E- ik-\delta+1))}\Gamma(-\delta)\,,\label{fourierscalar2pt}
\end{equation}
for $\delta \notin \mathbb{Z}^{*+}$ and
\begin{equation}
\begin{split}
    \mathcal{G}^\Delta(\omega_E,k)&\propto\frac{\Gamma(\frac{1}{2}(\omega_E+ ik+\delta+1))\Gamma(\frac{1}{2}(\omega_E- ik+\delta+1))}{\Gamma(\frac{1}{2}(\omega_E+ ik-\delta+1))\Gamma(\frac{1}{2}(\omega_E- ik-\delta+1))}\\
    &\qquad\times\left[\psi(\tfrac{1}{2}(\omega_E+ ik+\delta+1))+\psi(\tfrac{1}{2}(\omega_E- ik+\delta+1))\right]\,,\label{eq-Gscalar}
\end{split}
\end{equation}
for $\delta \in \mathbb{Z}^{*+}$.

First, let us investigate the pole-skipping points for $\delta \notin \mathbb{Z}^{*+}$. Note that the structure of \eqref{fourierscalar2pt} is of the form $\frac{\Gamma(x +\delta/2)}{\Gamma(x-\delta/2)}\frac{\Gamma(y +\delta/2)}{\Gamma(y-\delta/2)}$. Thus, if $\delta$ is a negative integer ($\delta\in\mathbb Z^-$) it boils down to the inverse of $|-\delta|$-th polynomials of $x$ and $y$:
\begin{equation} \label{plhyt}
    \frac{1}{\underbrace{(x-\delta/2-1)(x-\delta/2-2)\cdots(x+\delta/2)}_{|-\delta|\ \mathrm{factors}}} \cdot \frac{1}{x \to y} \,.
\end{equation}
Because this structure can give only poles, no pole-skipping point arises in this case.  In fact, the only possible negative integer value of $\delta$ is $-1$ because of the unitarity bound on the conformal dimension of the scalar field $\delta\geq-1$. Concretely, for $\delta=-1$, $\mathcal{G}^\Delta(\omega_E,k)\propto(\omega_E^2+k^2)^{-1}$, which has no pole-skipping point. 

Next, we move to the non-integer $\delta$ case ($\delta \notin \mathbb{Z}^{*+}$ and $\delta\notin\mathbb Z^-$).  We can see that the zeros come from the singular parts of gamma functions in the denominator and the poles come from the singular part of the gamma functions in the numerator. The pole-skipping points can be read from the intersections between zeros and poles of the two point function (\ref{fourierscalar2pt}). 

For the zeros and poles of the scalar field's two point function (\ref{fourierscalar2pt}), we obtain:
\begin{equation}\label{spss}
\begin{split}
	\text{Zeros} &: \omega_E\pm ik-\delta+1=-2i,\qquad(i=0,1,2\dots)\,,\\
	\text{Poles} &: \omega_E\pm ik+\delta+1=-2j,\qquad(j=0,1,2\dots)\,.
\end{split}
\end{equation}
These two sets of linear equations intersect at the points
\begin{equation} 
    \omega_{E,n}=-n, \ \ \mathrm{and}\ \ 
    ik_{n,q}=\pm\left(-n+2q+\delta-1\right)\label{sps}\,,
\end{equation}
where $n=1,2,\cdots$, $q=1,2,\cdots,n$, and  $\delta>-1$ because of the unitarity bound\footnote{$\delta=-1$ is excluded because of \eqref{plhyt}.} of the conformal dimension of the scalar field.  Specially,  $n=1$ contributes to the leading order, giving the leading pole-skipping points 
\begin{equation}
    \omega_{E*}=-1 \ \ \mathrm{and}\ \ 
    ik_*=\pm \delta \,.\label{lps}
\end{equation}

In order to visualize the pole-skipping structure, we make a plot of \eqref{fourierscalar2pt} when $d=4$ with $\Delta=4.5$ in Fig. \ref{fig:scalar}. The red lines are the lines of poles and the blue lines are the lines of zeros. Thus, the intersections of the red lines and blue lines are pole-skipping points, which are marked as white ``stars'' and circles. Among them, two white stars on the top are the leading pole-skipping points and the other white circles are the sub-leading pole-skipping points. The numerical values of these points agree with the \eqref{sps}.
Finally, for the case with $\delta \in \mathbb{Z}^{*+}$ considering the digamma function in \eqref{eq-Gscalar} {carefully~\cite{pspcomments} we} find that \eqref{sps} still holds.

\begin{figure}
\centering
    {\includegraphics[width=4.9cm]{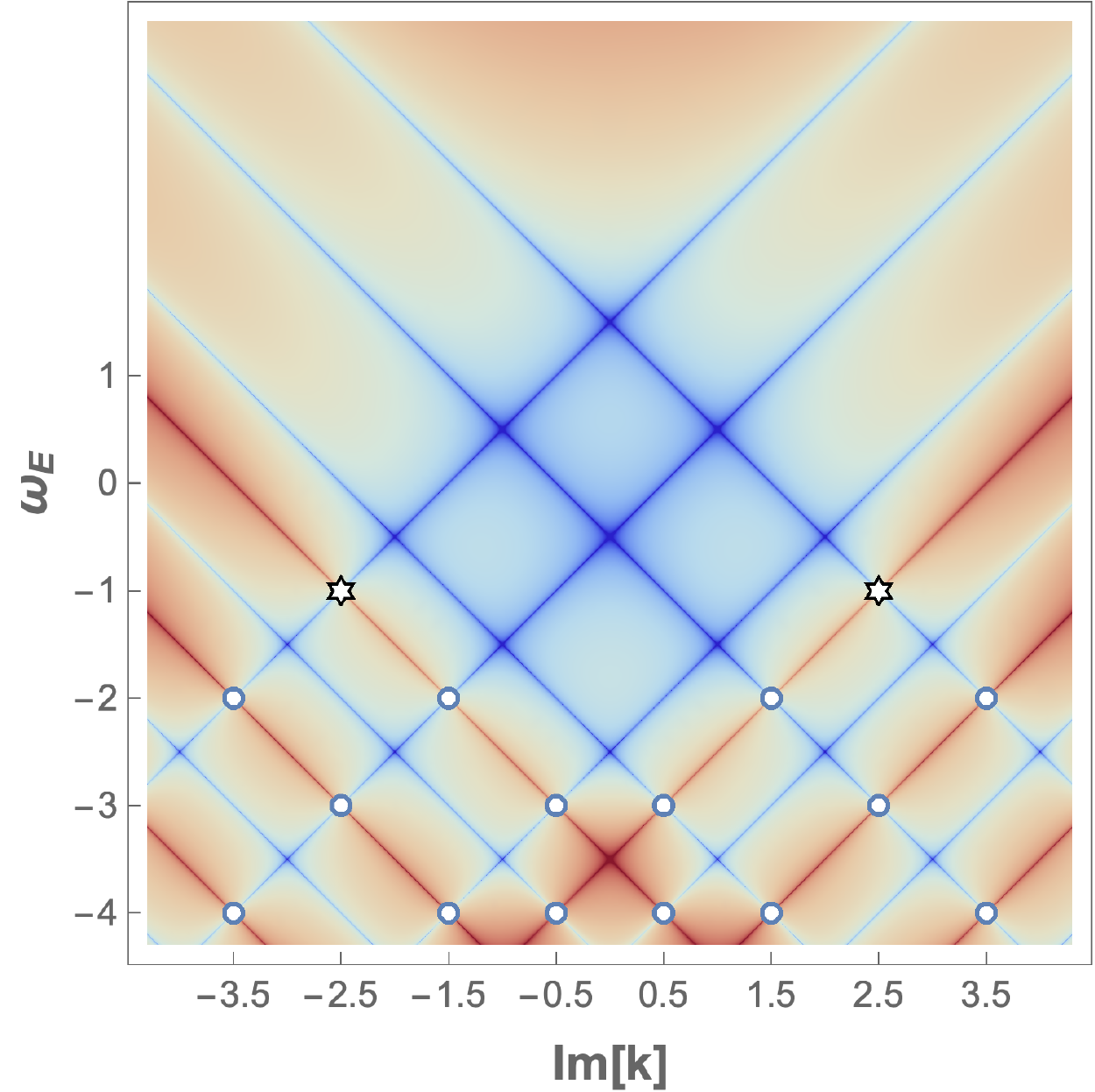}}
    \caption{$\log|\mathcal{G}^\Delta(\omega_E,k)|$: logarithm of the Fourier transformed conformal two point functions of scalar fields \eqref{fourierscalar2pt} for {$d=4,\,\Delta=4.5$.} The blue lines represent the zeros and the red lines represent the poles of the two point function so that the intersections between the red and blue lines are pole-skipping points, which are marked as white stars (leading points) and circles (sub-leading points). The numerical values of these points agree with the \eqref{sps}. }
    \label{fig:scalar}
\end{figure}

\subsection{Vector field}\label{tau2pt}
From the arguments of \cite{Costa:2011mg}, we can get the conformal two point function of the symmetric traceless spin $\ell$ primaries $\mathcal O^\Delta_{\mu_1\dots\mu_\ell}(P)$ with projection operator $\mathbb{P}^{A_1\dots A_\ell}_{\mu_1\dots\mu_\ell}(P)$ and auxiliary fields $Z^{A_i}$:
\begin{equation}
\begin{split}
& \mathcal G^\Delta_{\mu_1\dots \mu_\ell,\nu_1\dots \nu_\ell}(P_1,P_2):=\langle \mathcal O^\Delta_{\mu_1\dots\mu_\ell}(P_1)\mathcal O^\Delta_{\nu_1\dots\nu_\ell}(P_2)\rangle_{S^1\times\mathbb{H}^{d-1}}\\
\propto\; &\mathbb{P}^{A_1\dots A_\ell}_{\mu_1\dots\mu_\ell}(P_1)\mathbb{P}^{B_1\dots B_\ell}_{\nu_1\dots\nu_\ell}(P_2)\frac{\partial}{\partial Z^{A_1}_1}\dots\frac{\partial}{\partial Z^{A_\ell}_1}\frac{\partial}{\partial Z^{B_1}_2}\dots\frac{\partial}{\partial Z^{B_\ell}_2}\frac{H_{12}^\ell}{(-2P_1\cdot P_2)^{\Delta+\ell}}\,,
\end{split}
\end{equation}
where $H_{12}:= (P_1\cdot P_2)(Z_1\cdot Z_2)-(P_1\cdot Z_2)(P_2\cdot Z_1)$. For example, the two point function of the energy momentum tensors in momentum space was computed in \cite{Haehl:2019eae}. For the spin $1$ case, we can get the vector field's two point function as
\begin{equation}
\begin{split}
\mathcal{G}^\Delta_{\mu,\nu}(P_1,P_2)&:=\langle V^\Delta_{\mu}(P_1)V^\Delta_{\nu}(P_2)\rangle_{S^1\times\mathbb{H}^{d-1}}\\
&\propto\mathbb{P}^{A}_{\mu}(P_1)\mathbb{P}^{B}_{\nu}(P_2)\frac{\partial}{\partial Z^A_1}\frac{\partial}{\partial Z^B_2}\frac{(P_1\cdot P_2)(Z_1\cdot Z_2)-(P_1\cdot Z_2)(P_2\cdot Z_1)}{(-2P_1\cdot P_2)^{\Delta+1}}\\
&=\mathbb{P}^{A}_{\mu}(P_1)\mathbb{P}^{B}_{\nu}(P_2)\frac{(P_1\cdot P_2)\eta_{AB}-(P_{1B})(P_{2A})}{(-2P_1\cdot P_2)^{\Delta+1}}\,,\label{gauge2pt}
\end{split}
\end{equation}
where $V^\Delta_{\mu}(P)$ is a primary vector field with conformal dimension $\Delta$ and $\mathbb{P}^A_\mu(P)=\frac{\partial P^A}{\partial x_R^\mu}$ is the projection operator for spin 1 case of which $x_R^\mu=(\tau,\rho,x_\perp^i)$ is the Rindler coordinate.

\subsubsection{Longitudinal channel}
First, we consider the $\tau,\tau$ component of two point function, which we will call ``longitudinal channel''\footnote{Note that ``longitudinal channel'' does not mean a longitudinal space direction in hyperbolic space.} motivated by the holographic analysis in the following sections. We will  compare our results here with the pole-skipping analysis from holographic perspective in subsections \ref{retardscalarrmode} and \ref{horizonscalarmode}. From (\ref{gauge2pt}), we have 
\begin{equation}
\begin{split}
\mathcal{G}_{\tau,\tau}^\Delta(P_1,P_2)&\propto \mathbb{P}^A_\tau(P_1)\mathbb{P}^B_\tau(P_2)\frac{(P_1\cdot P_2)\eta_{AB}-(P_{1B})(P_{2A})}{(-2P_1\cdot P_2)^{\Delta+1}}\\
&=\mathcal{G}^{\Delta+1}(P_1,P_2)\left(1+\cos(\tau_1-\tau_2) Y_1\cdot Y_2\right).\label{gaugetau}
\end{split}
\end{equation}
To perform the Fourier transformation of \eqref{gaugetau}, we introduce the parameterized two point function as in \cite{Haehl:2019eae} which makes the Fourier transformation much easier than the direct computation:
\begin{equation}
\mathcal{G}_{(a,b)}^\Delta(P_1,P_2):=\frac{1}{(-2a\cos(\tau_{1}-\tau_2)-2bY_1\cdot Y_2)^\Delta}\,,\label{ptp}
\end{equation}
where we will take the $a,b\rightarrow1$ limit at the end. 
This parameterized scalar two point function is useful for the Fourier transformation of the two point functions because we can use
\begin{alignat}{1}
\cos(\tau_1-\tau_2)\;&\mathcal{G}^{\Delta}\rightarrow \frac{1}{2(\Delta-1)}\partial_a\mathcal{G}^{\Delta-1}_{(a,b)}\vert_{a,b\rightarrow1},\quad Y_1\cdot Y_2\;\mathcal{G}^{\Delta}\rightarrow\frac{1}{2(\Delta-1)}\partial_b\mathcal{G}^{\Delta-1}_{(a,b)}\vert_{a,b\rightarrow1}\,.\label{coord2der}
\end{alignat}

By using (\ref{ptp}) and (\ref{coord2der}), we can express the two point function as a sum of the parameterized scalar two point function differentiated by  parameters $a$ and $b$. Therefore, the Fourier transformation of such complicated expression only needs the single Fourier transformation of the parameterized scalar two point function {(see the details in appendix \ref{integration})}:
\be
    \mathcal{G}^\Delta_{(a,b)}(\omega_E,k)=\frac{\pi^{d/2}}{\Gamma(\Delta)}\frac{a^{\omega_E}}{b^{\omega_E+\Delta}}\frac{|\Gamma(\alpha)|^2}{\Gamma(\alpha+\alpha^*+\tfrac{d}{2}-\Delta)}\;_2F_1\left(\alpha,\alpha^*;\alpha+\alpha^*+\tfrac{d}{2}-\Delta;\frac{a^2}{b^2}\right)\,,
\ee
where $\alpha:= \frac{1}{2}(\omega_E+ik-\frac{d-2}{2}+\Delta)$.
Thus, we can compute the Fourier transformation of \eqref{gaugetau} as
\begin{alignat}{2}
\mathcal{G}_{\mathrm{longi}}^\Delta(\omega_E,k)=&    \left.\left[ \mathcal{G}^{\Delta+1}_{(a,b)}(\omega_E,k)+\frac{1}{4(\Delta-1)\Delta}\partial_a\partial_b \mathcal{G}^{\Delta-1}_{(a,b)}(\omega_E,k)\right] \right\vert_{a,b\rightarrow1}\,.\label{paramhyper}
\end{alignat}

Like the scalar field cases \eqref{fourierscalar2pt} and \eqref{eq-Gscalar}, the two point function for the longitudinal channel can be expressed in two ways depending on $\delta$: 
\begin{equation}
\begin{split}
    \mathcal{G}_{\mathrm{longi}}^\Delta(\omega_E,k)&\propto\frac{\Gamma(\frac{1}{2}(\omega_E+ ik+\delta))\Gamma(\frac{1}{2}(\omega_E- ik+\delta))}{\Gamma(\frac{1}{2}(\omega_E+ ik-\delta+2))\Gamma(\frac{1}{2}(\omega_E- ik-\delta+2))}\Gamma(-\delta)\\
&\qquad\times\left[\left(k^2+\delta^2\right)(\Delta-1)+\omega_E^2(d-\Delta-1)\right]\,,\label{fouriergauge2pt}
\end{split}
\end{equation}
for $\delta\notin\mathbb Z^*$  and 
\begin{equation}
\begin{split}
    \mathcal{G}_{\mathrm{longi}}^\Delta(\omega_E,k)&\propto\frac{\Gamma(\frac{1}{2}(\omega_E+ ik+\delta))\Gamma(\frac{1}{2}(\omega_E- ik+\delta))}{\Gamma(\frac{1}{2}(\omega_E+ ik-\delta+2))\Gamma(\frac{1}{2}(\omega_E- ik-\delta+2))}\\
&\qquad\times\left[\left(k^2+\delta^2\right)(\Delta-1)+\omega_E^2(d-\Delta-1)\right]\\
&\qquad\times\left[\psi(\tfrac{1}{2}(\omega_E+ ik-\tfrac{d}{2}+\Delta))+\psi(\tfrac{1}{2}(\omega_E- ik-\tfrac{d}{2}+\Delta))\right] 
\,,\label{rvtpf}
\end{split}
\end{equation}
for $\delta\in\mathbb Z^*$.

Let us first investigate the pole-skipping points for $\delta \notin \mathbb{Z}^{*+}$. Different from the scalar field case, $\delta$ cannot take negative integer values because the unitarity bound for the vector field, $\Delta \ge d-1$, and the condition, $d\geq3$, give
\begin{equation}
   {\delta\geq 1/2\quad\mathrm{and}\quad \Delta\geq2  \,.\label{vectorunitarity}}
\end{equation}
Therefore, the reduction of the gamma functions to \eqref{plhyt} does not happen when $\delta \notin \mathbb{Z}^{*+}$. We find zeros and poles from (\ref{fouriergauge2pt}) as 
\begin{alignat}{2}
	&\text{Zeros} &&: \left(k^2+\delta^2\right)(\Delta-1)+\omega_E^2(d-\Delta-1)=0,\label{case1}\\
	&\text{Zeros} &&:  \omega_E\pm ik-\delta+2=-2i\qquad\,\, i = 0,1,2\dots ,\label{case2}\\
	&\text{Poles} &&:   \omega_E\pm ik+\delta=-2j\qquad\qquad j=0,1,2\dots \label{case3}.
\end{alignat}
There are two possibilities for the intersection between zeros and poles: i) zeros from the polynomial (\ref{case1}) and ii) zeros from the linear equation (\ref{case2}). 

For the case i), {$\omega_E$ coming from the intersections between \eqref{case1} and the pole lines on the top (\eqref{case3} with $j=0$) is $0,1-\Delta$. With the condition \eqref{vectorunitarity},} the leading pole-skipping points are
\begin{equation}
	\omega_{E*}=0\ \ \mathrm{and}\ \  ik_*=\pm\delta\,.\label{pscft}
\end{equation}
For the case ii), there are sub-leading pole-skipping points:
\begin{equation}
    \omega_{E,n}=-n\ \ \mathrm{and}\ \ 
     ik_{n,q}=\pm\left(-n+2q+\delta-2\right)\,,\label{diffsub}
\end{equation}
where $n=1,2,\cdots$ and $q=1,2,\cdots,n$.
For the case with $\delta\in\mathbb Z^*$, it turns {out~\cite{pspcomments}} that the pole-skipping points   \eqref{pscft} and \eqref{diffsub} still hold. {Note that \eqref{vectorunitarity} excludes the case $\delta=0$.}

\subsubsection{Transverse channel}\label{transverse2pt} 
Next, we consider the two point function of vector fields with transverse components\footnote{For the $V_\tau$ component, we do not need to care about raising and lowering indices because $g_{\tau\tau}=1$.} $\langle V^\Delta_{ x^i_\perp}(P_1)V^{\Delta \,x^i_\perp}(P_2)\rangle_{S^1 \times \mathbb{H}^{d-1}}=\rho^2_2\,\langle V^\Delta_{ x^i_\perp}(P_1)V^{\Delta}_{x^i_\perp}(P_2)\rangle_{S^1 \times \mathbb{H}^{d-1}}$, which we will call ``transverse channel'' motivated by the holographic analysis in the following sections. We will  compare our results here with the pole-skipping analysis from the holographic perspective in subsections \ref{retardvectormode} and \ref{horizonvectormode}. We can continue from (\ref{gauge2pt}) to get the two point function of transverse components:
\begin{equation}
\begin{split}
\mathcal{G}_{x^i_\perp}^{\Delta \,x^i_\perp}(P_1,P_2)&=\langle V^\Delta_{x^i_\perp}(P_1)V^{\Delta \,x^i_\perp}(P_2)\rangle_{S^1 \times \mathbb{H}^{d-1}}\\
&\propto\;\rho^2_2\,\mathbb{P}^A_{x^i_\perp}(P_1)\mathbb{P}^B_{x^i_\perp}(P_2)\frac{(P_1\cdot P_2)\eta_{AB}-(P_{1B})(P_{2A})}{(-2P_1\cdot P_2)^{\Delta+1}}\\
&=\left(-\frac{\rho_2}{2\rho_1}\mathcal{G}^{\Delta}(P_1,P_2)+\frac{(x^i_{1\perp}-x^i_{2\perp})^2}{\rho_1^2}\mathcal{G}^{\Delta+1}(P_1,P_2)\right).
\label{gauge2ptvector}
\end{split}
\end{equation}

Unlike the longitudinal channel, we cannot use the differentiation trick \eqref{coord2der}. Indeed, for the longitudinal channel, which is relevant with $V^\tau$, we may use an eigenfunction $f$ \eqref{eigen1} of the scalar Laplacian $\square_{S^1\times\mathbb{H}^{d-1}}$ \eqref{eesc} because $D^\mu D_\mu V^\tau= \square_{S^1\times\mathbb{H}^{d-1}} V^\tau$.\footnote{Here, $\mu=({\tau,\rho,\vec x_\perp})$ and $D_\mu$ is a covariant derivative on $S^1\times\mathbb{H}^{d-1}$.} 
However, this is not the case for the transverse component, i.e.
$ D^\mu D_\mu V^j\neq \square_{S^1\times\mathbb{H}^{d-1}} V^j
$
for the remaining vector field's components $j=(\rho,x_\perp^1,\dots, x_\perp^{d-2})$. See for example \eqref{eevc}. 

Thus, we need to find the ``Fourier mode'' for the transverse channel. For this goal, let us first recall~\cite{Ueda:2018xvl} that the transverse component of the vector field can be expressed in terms of the vector harmonics\footnote{The eigenfunctions $f(P;\omega_E, k,\vec p_\perp)$ in the previous section are called scalar harmonics~\cite{Ueda:2018xvl}.} $\mathfrak f_j(P;p)$ in $\mathbb{H}^{d-1}$:
\begin{equation}
V_j(P)=\sum_p v(p)\, \mathfrak f_j(P;p) \,, \qquad D^j\mathfrak f_j(P;p)=0\,, \label{vectorharmon}
\end{equation}
where $\mathfrak f_{x^j_\perp}(P;p)$ is a shorthand notation for $\mathfrak f_{x^j_\perp}(P;\omega_E, k,\vec p_\perp)$. 
For more details of the decomposition of the vector field, we refere to \cite{Ueda:2018xvl} or section \ref{retardvectorfield}.

The vector harmonics $\mathfrak f_{x^i_\perp}(P;p)$ play the role of a ``Fourier mode'' so we need to compute the eigenfunction of the following differential operator
\begin{align}
D^\mu D_\mu\mathfrak f_{x^i_\perp}(P;p)= [\partial_\tau^2+\rho^2\partial_\rho^2-(d-5)\rho\partial_\rho +\rho^2 \square_{\mathbb{R}^{d-2}}-(d-2)]\mathfrak  f_{x^i_\perp}(P;p)\,.\label{eevc}
\end{align}
Note that this may be related with the Laplacian for scalar fields \eqref{eesc}
\begin{align*}
\square_{S^1\times\mathbb{H}^{d-1}}=\partial_\tau^2+\rho^2\partial_\rho^2-(d-3)\rho\partial_\rho +\rho^2 \square_{\mathbb{R}^{d-2}},
\end{align*}
in the sense that the coefficient of $\rho\partial_\rho$ in (\ref{eevc}) is related with the one in (\ref{eesc}) with the replacement $d\to d-2$.  Therefore, the eigenfunction $\mathfrak f_{x^i_\perp}(P;p)$ can be obtained from $f(P;p)$ in (\ref{eigen1}) with the replacement $\rho^{\frac{d-2}{2}}\to \rho^{\frac{d-4}{2}}$ :
\begin{equation}
\begin{split}
f(P;p)&\ \propto \ \rho^{\frac{d-2}{2}}K_{ik}(|p_\perp|\rho)e^{i(\omega_E\tau+\vec p_\perp\cdot\vec  x_\perp)}\,,\\
\rightarrow\quad\mathfrak f_{x^i_\perp}(P;p)&\ \propto \ \rho^{\frac{d-4}{2}}K_{ik}(|p_\perp|\rho)e^{i(\omega_E\tau+\vec p_\perp\cdot\vec  x_\perp)}\ \propto \ \frac{f(P;p)}\rho \,,\label{vectoreigen}
\end{split}
\end{equation}
with the eigenvalue
\be
    -\omega_E^2-k^2-\left(\frac{d-4}{2}\right)^2-(d-2)\,.\label{transeigen}
\ee
The replacement $d\to d-2$ between  eigenfunctions is consistent with eigenmodes (\ref{ssh}) and (\ref{vsh}) in the holographic computation.}
The functions $\mathfrak f_{x^j_\perp}(P;p)$ should satisfy the divergenceless condition in \eqref{vectorharmon} so we impose 
\begin{equation}
p^j_\perp=0 \,, \quad \mathrm{if}\ \ j = i \,,
\end{equation}
for the specific $i$-th mode, $\mathfrak f_{x^i_\perp}(P;p)$.

Let us ``Fourier-transform'' the two point function of the transverse channel of the vector field by using the eigenfunction $\mathfrak f_{x^i_\perp}(P;p)$:
\begin{align}
    &\mathcal{G}^{\Delta}_{\mathrm{transv}}(\omega_E,k)\,\mathfrak{f}_{x^i_\perp}(P)=\int dP'\, \mathcal{G}_{x^i_\perp}^{\Delta \,x^i_\perp}(P,P')\,\mathfrak{f}_{x^i_\perp}(P')\,.
\end{align}
By plugging the expressions \eqref{gauge2ptvector} and \eqref{vectoreigen}:
\begin{align}
    \mathcal{G}^{\Delta}_{\mathrm{transv}}(\omega_E,k)\,\frac{f(P;p)}{\rho}&=\int dP'\, \left(-\frac{\rho'}{2\rho}\mathcal{G}^{\Delta}(P,P')+\frac{(x^i_{\perp}-x'^i_{\perp})^2}{\rho^2}\mathcal{G}^{\Delta+1}(P,P')\right)\,\frac{f(P')}{\rho'}\nonumber\\
    &=\int dP'\, \left(-\frac{1}{2}\mathcal{G}^{\Delta}(P,P')+\frac{(x^i_{\perp}-x'^i_{\perp})^2}{\rho\rho'}\mathcal{G}^{\Delta+1}(P,P')\right)\,\frac{f(P')}{\rho}\nonumber\\
    &=\left(-\frac{1}{2}+\frac1{2\Delta}\right)\mathcal{G}^{\Delta}(\omega_E,k)\frac{f(P;p)}\rho\,,
\end{align}
where $\mathcal{G}^{\Delta}(\omega_E,k)$ is the Fourier transformed scalar field's two point function \eqref{fttf11}. In the computation, we used the Gaussian integration: 
\begin{equation}
    \int d^{d-2}\vec x'_\perp(x_{\perp}^i-x_{\perp}^{\prime i})^2e^{-\frac{(\vec x_{\perp}-\vec x'_{\perp})^2}{2\rho\rho'}\zeta+i\vec p_\perp\cdot (\vec x'_\perp-\vec x_\perp)}\stackrel{p^i_\perp\rightarrow0}{=}e^{-\frac{\rho\rho'}{2\zeta}|\vec p_\perp|^2}\left(\frac{2\pi\rho\rho'}{\zeta}\right)^{\frac{d-2}{2}}\frac{\rho\rho'}{\zeta}\,,
\end{equation}
and 
\begin{equation}
\begin{split}
    &\int dP'\, \frac{(x^i_{\perp}-x'^i_{\perp})^2}{\rho\rho'}\mathcal{G}^{\Delta+1}(P,P')\,\frac{f(P')}{\rho}\\
    &=\frac{(2\pi)^{\frac{d}{2}}\rho^{\frac{d-4}{2}}}{2^{\Delta+1}\Gamma(\Delta+1)}\int_0^\infty d\zeta\,\zeta^{\delta}\,I_{\omega_E}(\zeta)\int_0^\infty \frac{d\rho'}{\rho'}\,e^{-\frac{\rho^2+\rho'^2}{2\rho\rho'}\zeta}K_{ik}(|p_\perp|\rho')e^{-\frac{\rho\rho'}{2\zeta}|\vec p_\perp|^2}\\
    &=\frac{1}{2\Delta}\mathcal{G}^{\Delta}(\omega_E,k)\,,
\end{split}
\end{equation}
where the same techniques are applied as in \eqref{longcomp}. Then the result becomes
\begin{alignat}{2}
\mathcal{G}_{\mathrm{transv}}^{\Delta}&(\omega_E,k)=-\frac{1}{2}\left(\frac{\Delta-1}{\Delta}\right)\mathcal{G}^{\Delta}(\omega_E,k)\label{transversehyper}\,.
\end{alignat}
 Note that \eqref{transversehyper} is equivalent to the Fourier transformed scalar two point function \eqref{fttf11} multiplied by $\frac{\Delta-1}\Delta$.\footnote{{However, unlike the scalar field's two point function, \eqref{vectorunitarity} restricts the case when $\delta$ is zero or negative integer values. \eqref{vectorunitarity} also guarantees that $\frac{\Delta-1}{\Delta}$ is regular.} Thus, we can ignore this factor since it does not affect the pole-skipping structure. } Thus, \eqref{transversehyper} gives the results
\begin{align}
    \mathcal{G}_{\mathrm{transv}}^{\Delta}(\omega_E,k)\propto&\frac{\Gamma(\frac{1}{2}(\omega_E+ ik+\delta+1))\Gamma(\frac{1}{2}(\omega_E- ik+\delta+1))}{\Gamma(\frac{1}{2}(\omega_E+ ik-\delta+1))\Gamma(\frac{1}{2}(\omega_E- ik-\delta+1))}\Gamma(-\delta)\,,\label{fouriergauge2ptvector0}
\end{align}
for $\delta \notin \mathbb{Z}^{*+}$ and
\begin{equation}
\begin{split}
    \mathcal{G}_{\mathrm{transv}}^{\Delta}(\omega_E,k)\propto&\frac{\Gamma(\frac{1}{2}(\omega_E+ ik+\delta+1))\Gamma(\frac{1}{2}(\omega_E- ik+\delta+1))}{\Gamma(\frac{1}{2}(\omega_E+ ik-\delta+1))\Gamma(\frac{1}{2}(\omega_E- ik-\delta+1))}\\
    &\qquad\times\left[\psi(\tfrac{1}{2}(\omega_E+ ik+\delta+1))+\psi(\tfrac{1}{2}(\omega_E- ik+\delta+1))\right]\,,
\end{split}
\end{equation}
for $\delta \in \mathbb{Z}^{*+}$.
 Therefore, the pole-skipping structure of \eqref{transversehyper} is same as the scalar two point function's \eqref{sps}:\footnote{We do not compute the pole-skipping structure of the transverse channel with the $\rho$-component because of technical complications. Instead, we will show that the answer remains the same for all the transverse components in the holographic calculation in subsection \ref{retardvectormode}.} 
\begin{equation}
\label{scalarpspcft}
    \omega_{E,n}=-n \ \ \mathrm{and}\ \ 
    ik_{n,q}=\pm\left(-n+2q+\delta-1\right)\,,
\end{equation}
where $n=1,2,\cdots$ and $q=1,2,\cdots,n$ and the leading pole-skipping points $(n=1)$ are:
\begin{equation}
    \omega_{E*}=-1 \ \ \mathrm{and}\ \ 
    ik_*=\pm\delta\,,\label{slvpsp}
\end{equation}
with the different condition $\delta\geq1/2$ ($\delta>-1$ for scalar field).

\subsection{Two exponents \texorpdfstring{$\omega$}{w} and \texorpdfstring{$L$}{L} from the pole-skipping points}

In the previous subsections, we computed the pole-skipping points of two point functions. Interestingly, the leading pole-skipping points turn out to appear also in the four point OTOC functions in the late time and large distance limit, as interesting physical observables. Typical examples are the Lyapunov exponent and butterfly velocity in the case of the energy-momentum tensor operator.

It was also argued in ~\cite{Haehl:2019eae} that this relation between two point functions and OTOC  can be seen in the basic “Fourier” modes \eqref{eigen1} in the late time and large distance limit. Let us start with the complex conjugate of \eqref{eigen1}
\begin{align}
    f^*&\propto e^{-i \omega_E\tau } \rho^{\frac{d-2}{2}}J^*_{ik}(|p_\perp|\rho) \\
    & \sim e^{ \omega_E t } \rho^{\frac{d-2}{2}-ik}  \quad (\rho \to 0) \\
    & \sim e^{ \omega_E t } e^{\left(-\frac{d-2}{2}+ik\right) \mathbf{d}} 
    \,,
    \label{eigen111}
\end{align}
where we drop the transverse direction ($\vec{x}_\perp=0$) for simplicity  and used $J^*_{ik}(i|p_\perp|\rho)$ instead of $K^*_{ik}(|p_\perp|\rho)$ because the former is regular for imaginary $k$.\footnote{If $ik$ is a positive non-integer, $J^*_{ik}(i|p_\perp|\rho)$ is not regular at $\rho\to0$. In this case, we still use $J^*_{ik}(i|p_\perp|\rho)$ for the purpose of comparison with OTOC.}
(Note that we are interested in the leading pole-skipping points, where $k$ is imaginary.) In the second line we used the analytic continuation $\tau = it$ and the  approximation $J^*_{ik}(i|p_\perp|\rho) \to \rho^{-ik}$ as $\rho \to 0$, which means the large distance limit as shown below. In the last line we used 
\begin{align}
e^{\mathbf{d}} = e^{\mathbf{d}(1,3)}\propto \frac{1}{\rho_{3}} \;\;\;\;\;(\rho_3\to0 \ \ \mathrm{at} \ \ \mathrm{fixed} \ \ \rho_1)\,,
\end{align}
which can be derived from \eqref{dd} in the large distance limit, $\mathbf{d} \gg 1$.

In the previous subsections, we found the leading pole-skipping points $(\omega_{E*}, k_*)=(\ell-1, \pm i(\Delta-d/2))$ for scalar fields $(\ell=0)$ and vector fields $(\ell=1)$. At these leading pole-skipping points, \eqref{eigen111} becomes
\begin{equation}
    f^* \sim e^{(\ell-1)t + \left(-\frac{d-2}{2}\mp (\Delta-\frac{d}{2})\right)\mathbf{d}} \,. \label{comp1234}
\end{equation}
Comparing this with the form $e^{-i\omega_* + L_* \mathbf{d}}$ in \eqref{dol}, we find
\begin{align}
\omega_*=i(\ell-1)\,, \quad L^\pm_*=-\frac{d-2}{2}\mp \left(\Delta-\frac{d}{2}\right) = 1-\Delta \ \ \mathrm{or} \ \ \Delta - d +1\,,
\end{align}
which agree with (\ref{eol}) and (\ref{eols}).

For the transverse channel of the vector field, the basis ``Fourier'' mode is $\mathfrak f_{x^i_\perp}$ in \eqref{vectoreigen}. By the same procedure with the pole-skipping point (\ref{slvpsp})
we obtain
\begin{align}
\omega_* = -i\,, \quad L^\pm_*=-\frac{d-4}{2}+ik=-\frac{d-4}{2}\mp \left(\Delta-\frac{d}{2}\right) = 2-\Delta \ \ \mathrm{or} \ \ \Delta - d +2\,.
\end{align} 
However, in the vector sector, this is sub-leading because the longitudinal channel is leading.

\section{Pole-skipping analysis: bulk retarded Green's functions\label{SHGF}}
In this section, we derive the real-time retarded Green's function of scalar and vector fields for a Rindler-AdS$_{d+1}$ geometry
and compute the corresponding pole-skipping points. Consider the Rindler-AdS$_{d+1}$ geometry, with metric given by
\be \label{eq-metric}
ds^2=-\left(\frac{r^2}{\ell_\mathrm{AdS}^2}-1\right)dt^2+\frac{dr^2}{\frac{r^2}{\ell_\mathrm{AdS}^2}-1}+r^2 dH_{d-1}^2,
\ee
where $dH_{d-1}^2=d\chi^2+\sinh^2\chi \, d\Omega_{d-2}^2$ is the line element (squared) of the $(d-1)$-dimensional hyperbolic space $\mathbb{H}^{d-1}$, and $d\Omega_{d-2}$ is the line element of a unit sphere $S^{d-2}$. The Hawking temperature is $ T = 1/(2\pi \ell_{\text{AdS}})$. From here, we set the AdS radius to unity. 
It is convenient to introduce a new radial coordinate defined by $r=\cosh \nr  $, in terms of which the metric becomes\footnote{The coordinate patch is different from \eqref{cftco}.}
\be \label{rhoco}
ds^2=-\sinh^2\nr \, dt^2+d\nr^2+\cosh^2\nr\, dH_{d-1}^2\,.
\ee

\subsection{Scalar field}\label{sec:scar}
We consider a minimally coupled scalar field, with action
\be
S_\mt{scalar}=-\frac{1}{2}\int d^{d+1}x \sqrt{-g} \left(g^{\mu \nu} \partial_{\mu} \phi \partial_{\nu} \phi+m^2 \phi^2 \right)\label{saction}\,,
\ee
propagating on the background (\ref{eq-metric}). The corresponding equation of motion is
\be
\frac{1}{\sqrt{-g}}\partial_{\mu} (\sqrt{-g}g^{\mu \nu} \partial_{\nu} \phi)-m^2 \phi=0\,.
\ee
In terms of the coordinates $(t,\nr,x^i)$, where $x^i \in \mathbb{H}^{d-1}$, this equation of motion can be written as
\be \label{eq-scalar}
\partial_{\nr}^2 \phi -\frac{\partial_t^2 \phi}{\sinh^2 \nr}+\frac{\square_{H^{d-1}} \phi }{\cosh^2 \nr}+ \big[ \coth \nr+(d-1) \tanh \nr \big] \partial_{\nr} \phi-m^2 \phi=0\,,
\ee
where $\square_{H^{d-1}}= \partial_{\chi}^2+(d-2) \coth \chi \partial_{\chi}+\frac{1}{\sinh^2\chi} \square_{S^{d-2}}$ is the Laplacian operator in $\mathbb{H}^{d-1}$.
To solve the above equation, we use the following ansatz
\be
\phi(t,\nr,x^i)= \sum_{L,M}F(t, \nr)Y_{L\,M}^{(d-1)}(i \chi, \Omega_i)\label{ssh}\,,
\ee
where $\Omega_i \in S^{d-2}$, and the hyperspherical harmonics $Y_{L\,M}^{(d-1)}(i \chi, \Omega_i)$ satisfy the equation\footnote{See Appendix \ref{ap-hyper} for more details about the hyperspherical harmonics $Y_{L\,M}^{(d-1)}$.}
\be
\square_{H_{d-1}} Y_{L\,M}^{(d-1)} (i \chi, \Omega_{i})= L(L+d-2) Y_{L\,M}^{(d-1)} (i \chi, \Omega_{i})\,.\label{YLM}
\ee
Plugging \eqref{ssh} and \eqref{YLM} into \eqref{eq-scalar} with $F(t, \nr) = \int \dd \omega e^{-i\omega t}F(\omega, \nr)$, we have
\be \label{eq-F}
F''(\nr)+\big[ \coth \nr+(d-1) \tanh \nr \big] F'(\nr)+\left[\frac{\omega^2}{\sinh^2\nr}+\frac{L(L+d-2)}{\cosh^2\nr}-m^2 \right]F(\nr)=0\,,
\ee
where the primes denote derivatives with respect to $\nr$ and we replace $F(\omega, \nr)\rightarrow F(\nr)$ for notational simplicity.

It is customary to express the solutions of (\ref{eq-F}) in terms of $z=\tanh^2\nr$. In this coordinate, the horizon is located at $z=0$ while the boundary is located at $z=1$. The incoming solution is given by
\begin{align}
\label{scalarsol}
&F_\mt{in}(z)=(1-z)^{\Delta_{+}/2}\,z^{-i \omega /2} {}_{2}F_1\left(a, b, a + b + \mathcal{N}; z \right)\,,\nonumber\\
   &a = \,\frac{1}{2}\left(-i \omega-L-(d-2-\Delta_{+})\right), \quad b = \frac{1}{2}\left(-i \omega+L+\Delta_{+}\right), \quad \mathcal{N} = \frac{d}{2}-\Delta_{+}\,,
\end{align}
where $\Delta_{+} =d/2+\sqrt{(d/2)^2+m^2}$. The outgoing solution can be obtained from $F_\mt{in}(z)$ by replacing $\omega \rightarrow - \omega$.

To compute the retarded Green's function, we need to know the asymptotic forms of the hypergeometric function near $z=1$, which are summarized in appendix \ref{holoregul}. 
By using \eqref{bulkFP} with $p=0$ we obtain the near-boundary behavior of \eqref{scalarsol}:
\begin{equation}
    F_\mt{in}(z)\approx(1-z)^{\frac{\Delta_{-}}{2}}A(\omega, L) + (1-z)^{\frac{\Delta_{+}}{2}}\Big(B(\omega, L)+C(\omega, L) \log(1-z) \Big)\,,
    \label{boundaryscar}
\end{equation}
where $\Delta_{-} = d - \Delta_{+} = 2\mathcal{N} + \Delta_{+}$. Since the coefficients $A$, $B$, and $C$ depend on whether $\mathcal{N}$ is integer or not,\footnote{$B$ contains digamma functions and $C\neq 0$ when $\mathcal{N}$ is a negative integer.} we consider two cases separately: $\mathcal{N}\in \mathbb{Z}^{*-} := \{ 0 \} \cup \mathbb{Z}^-$ or  $\mathcal{N}\notin \mathbb{Z}^{*-}$. Note that $\mathcal{N} = -\sqrt{(d/2)^2 + m^2}$ cannot be positive.

\subsubsection*{Non-integer $\mathcal{N} = \frac{d}{2} - \Delta_{+}$}
In $\mathcal{N}\neq \mathbb{Z}^{*-}$ case, the factors $A(\omega,L)$ and $B(\omega,L)$ in \eqref{boundaryscar} can be read off by using \eqref{nIcoef}. In the standard quantization, the conformal dimension $\Delta$ of the operator is identified with  $\Delta_{+}$ so the retarded Green's function is
\begin{align} \label{G123}
&G^R(\omega, L)\nonumber\\
&\propto \frac{B(\omega, L)}{A(\omega, L)}\propto \frac{\Gamma \left(+\frac{d}{2}-\Delta \right)}{\Gamma \left(-\frac{d}{2}+\Delta\right)} \frac{\Gamma \left( \frac{1}{2}(-i \omega-L-(d-2-\Delta)\right) \Gamma \left(\frac{1}{2}(-i \omega+L+\Delta) \right)}{\Gamma \left( \frac{1}{2}(- i \omega-L+2-\Delta)\right) \Gamma \left( \frac{1}{2}(-i \omega+L+d-\Delta) \right)}\,,
\end{align}
for non-integer $\Delta\geq d/2$.

To deal with the case $\Delta < d/2$ we may consider 
the alternative quantization, which identifies $\Delta_{-}$ with the conformal dimension $\Delta$. In this case, the meaning of the source and the response are interchanged so $G^R \propto \frac{A(\omega,L)}{B(\omega,L)}$. However, the final result remains the same as \eqref{G123}. An easy way to see is taking the inverse of \eqref{G123} and replace $\Delta \rightarrow d-\Delta$. See
\eqref{rels}, \eqref{nIcoefalter}, and \eqref{nIalter} for more details.

To compare this with the field theory computation \eqref{fourierscalar2pt} we replace $(\omega, L)$ with $(\omega_E, k)$ by 
\begin{equation} \label{replace123}
    \left(\omega, L \right)  \rightarrow \left(i \omega_E\,, \  \pm i k-\frac{d-2}{2} \right) \,,
\end{equation}
where the relation between $L$ and $\pm k$ (the sign does not matter) are obtained by the coordinate transformation i.e. by matching the eigenvalues \eqref{eigen1} and \eqref{YLM} in the two coordinate systems: $L(L+d-2) = -k^2-\frac{(d-2)^2}{4}.$ After the replacement, the retarded Green's function \eqref{G123} becomes
\be \label{G123p}
G^R(\omega, L) \to \frac{\Gamma(\frac{d}{2}-\Delta)}{\Gamma(-\frac{d}{2}+\Delta)}\frac{\Gamma(\frac{1}{2}(\omega_E+ ik-\frac{d-2}{2}+\Delta))\Gamma(\frac{1}{2}(\omega_E- ik-\frac{d-2}{2}+\Delta))}{\Gamma(\frac{1}{2}(\omega_E+ ik+\frac{d+2}{2}-\Delta))\Gamma(\frac{1}{2}(\omega_E- ik+\frac{d+2}{2}-\Delta))} \,,
\ee
which agrees with $\mathcal{G}^\Delta(\omega_E,k)$ in  \eqref{fourierscalar2pt} up to an unimportant numerical factor   
as far as the pole-skipping points are concerned.

The pole-skipping occurs at special values of $(\omega, L)$ such that the poles of the Gamma functions in the denominator and numerator coincide. We find the special frequencies and special values of $L$:
\bea
\label{scalarpsp}
\omega_n = - i\,n \ \ \mathrm{and}\ \ L_{n,q}=-\frac{d-2}{2}\pm\left(-n+2q+\Delta-\frac{d+2}{2}\right)\,, 
\eea
where $n=1,2,\cdots$ and $q=1,2,\cdots,n$.  More explicitly, $L_{n,q}=-n+2q+\Delta-d$ and $L_{n,q}=n-2q-\Delta+2$. The expression \eqref{scalarpsp} is more convenient to compare with the field theory result \eqref{sps}.
The first instance of pole skipping occurs for\footnote{Remember that, for our geometry, $2\pi T =1$.} $\omega_* = -i$ and $L^+_*=1-\Delta$ or $L^-_*=\Delta-d+1$. 

\subsubsection*{Integer $\mathcal{N} = \frac{d}{2} - \Delta_{+}$}
\label{posintescar}
In the $\mathcal{N}\in \mathbb{Z}^{*+}$ case, the boundary expansion of hypergeometric function in \eqref{scalarsol} also involves logarithmic terms, which are related to the matter conformal anomaly. 
In this case, we use \eqref{Icoef} to obtain $A$ and $B$ in \eqref{boundaryscar}. 
Then the retard Green's function for $\Delta = \Delta_{+}$ is given by 
\begin{align}
G^R(\omega, L) \propto \frac{B(\omega, L)}{A(\omega, L)} \propto & \frac{\Gamma \left( \frac{1}{2}(-i \omega-L-(d-2-\Delta)\right) \Gamma \left(\frac{1}{2}(-i \omega+L+\Delta) \right)}{\Gamma \left( \frac{1}{2}(- i \omega-L+2-\Delta)\right) \Gamma \left( \frac{1}{2}(-i \omega+L+d-\Delta) \right)} \times  \nonumber \\
&\left[ \psi\left( \frac{2+\Delta-L-d-i \omega}{2}\right)+\psi\left( \frac{\Delta+L-i \omega}{2}\right)\right]\label{jhg123hol}\,,
\end{align}
up to contact terms. The above result becomes subtle when $\mathcal{N}=0$. If $\mathcal{N}=0$, there is no distinction between $A$ and $B$, since $\Delta_{-} = \Delta_{+}$ in equation \eqref{boundaryscar}. In this case, $C$ plays the role of $A$ and the retarded Green's function can be obtained by using \eqref{zerocase}:
\begin{equation}
    G^R(\omega, L) \propto \frac{B(\omega, L)}{C(\omega, L)} \propto \psi\left(\frac{1}{2}\left(-i \omega-L-(d-2-\Delta_{+})\right)\right)+ \psi\left(\frac{1}{2}\left(-i \omega+L+\Delta_{+}\right)\right)\,.
    \label{scarzero}
\end{equation}
Note that \eqref{jhg123hol} and \eqref{scarzero} can also be obtained from \eqref{G123p} by using the prescription of replacing $\Gamma(\mathcal{N})/\Gamma(-\mathcal{N})$ with $\psi(a) + \psi(b)$.\footnote{We checked that the prescription works when the fields takes the form of \eqref{bulkFP}. See appendix \ref{holoregul} for more details.}

To check consistency with the field theory result, one can rewrite \eqref{jhg123hol} in terms of $(\omega_E, k)$ by using \eqref{replace123}:
\begin{align}
G^R(\omega_E, k) \propto &\frac{\Gamma(\frac{1}{2}(\omega_E+ ik-\frac{d-2}{2}+\Delta))\Gamma(\frac{1}{2}(\omega_E- ik-\frac{d-2}{2}+\Delta))}{\Gamma(\frac{1}{2}(\omega_E+ ik+\frac{d+2}{2}-\Delta))\Gamma(\frac{1}{2}(\omega_E- ik+\frac{d+2}{2}-\Delta))} \times \nonumber\\
&\left[ \psi\left(  \frac{1}{2}(\omega_E+ ik-\frac{d-2}{2}+\Delta) \right)+\psi\left( \frac{1}{2}(\omega_E- ik-\frac{d-2}{2}+\Delta)  \right)\right]\,, \label{jhg123}
\end{align}
which matches the corresponding field theory results (\ref{eq-Gscalar}).

\subsection{Vector field}\label{retardvectorfield}
In this section, we consider a minimally coupled massive vector field, with action
\begin{align}
\begin{split}
&S_{A} = -\int d^{d+1}x \sqrt{-g}\left( \frac{1}{4} F^2 +  \frac{1}{2}m^{2} A^2\right)\,,\label{paction}
\end{split}
\end{align}
propagating on the background (\ref{eq-metric}). The corresponding equations of motion are
\begin{equation}
\label{eqproca}
\nabla_{M}F^{M N} - m^2 A^{N} = 0\,.
\end{equation}
Let us consider the metric \eqref{rhoco},
\begin{align}
\label{metricdecom}
\dd s^2&=-\sinh^2\nr \, \dd t^2+d\nr^2+\cosh^2\nr\, \dd H_{d-1}^2 =: g_{ab}\dd y^a \dd y^b + u(y)^2 \gamma_{ij}\dd x^i \dd x^j,
\end{align}
where $y^1 = t$, $y^2=\nr\,$, $\gamma_{ij} \dd x^i \dd x^j = \dd H_{d-1}^2$ and 
\begin{equation} \label{eq-u}
  u(y) = \cosh \nr  \,.
\end{equation} 
In the hyperbolic space, a general perturbation of the dual vector field $A_\mu$ can be decomposed into ``longitudinal channel'' $(A^L_a,A^L)$ and ``transverse channel'' $A_i^T$ as follows \cite{Ueda:2018xvl}:
\begin{equation}
\label{decomvec}
A_{\mu}dx^{\mu} = A^L_a dy^a + \hat{D}_iA^Ldx^i + A_i^T dx^i, \quad \hat{D}^iA_i^T = 0\,,
\end{equation}
where the differential operators $\hat{D}_i$ denote covariant derivatives with respect to  $\gamma_{ij}$. 
Since the ``longitudinal channel'' $(A^L_a,A^L)$ and the ``transverse channel'' $A_i^T$ are independent of each other, we consider them separately one by one.

\subsubsection{Longitudinal channel}\label{retardscalarrmode} 
First, we derive the retarded Green's function corresponding to the massless vector field $A_t$, which belongs to the ``longitudinal channel", by using master field variables \cite{Ueda:2018xvl, Kodama:2003kk, Kodama:2003jz, Kodama:2000fa}. 
Even though this classification is still valid for the massive case~\cite{Ueda:2018xvl}, the advantage of the massless case is that the ``longitudinal channel" can be described by a {\it single} master field variable \cite{Kodama:2003kk}, which makes the computations tractable. For the massive case, since there is no gauge symmetry, we should consider three `scalar-type components'~\cite{Ueda:2018xvl}, which are coupled. Due to this technical difficulty we will consider a specific case, where $\omega=0$. This is not most general but general enough for our main purpose, which is finding the {\it leading} pole-skipping point.

\paragraph{Massless case}
For the ``longitudinal channel'', we choose the following form of perturbation:
\begin{equation}
\label{decomvecdif}
A_{\mu}dx^{\mu} = A^L_a dy^a + \hat{D}_iA^Ldx^i.
\end{equation}
In particular, the ``longitudinal channel" can be described in terms of the master variable $\mathcal{A}^L$, and scalar harmonics $\mathbb{S}_{k_S}$:
\begin{equation}
\label{difgauge}
\begin{split}
A^L_a &= \sum_{k_S}\left(D_a \mathcal{F}_{k_S}(y)+\frac{1}{u^{d-3}}\epsilon_{ab}D^b\mathcal{A}_{k_S}^L(y)\right)\mathbb{S}_{k_S}(x),\\
\hat{D}_i A^L &= \sum_{k_S} \mathcal{F}_{k_S}(y)\hat{D}_i\mathbb{S}_{k_S}(x),
\end{split}
\end{equation}
where $\epsilon_{ab}:= \sqrt{-\det(g_{ab})}\, \tilde{\epsilon}_{ab}$ is the Levi-Civita tensor with {$\tilde{\epsilon}_{12} = 1$} and $D_a$ is a covariant derivative with respect to $g_{ab}$ in \eqref{metricdecom}. {$u = \cosh \nr$ and $\hat{D}_i$ is covariant derivative with respect to $\gamma_{ij}$ in \eqref{metricdecom}.} Here, we newly introduced a gauge freedom $\mathcal{F}_{k_S}$, which is absent in \cite{Kodama:2003kk}. This is because we need to work  with the gauge field in the standard holographic scheme, while it is enough to consider the field strength in \cite{Kodama:2003kk}.
The scalar harmonics $\mathbb{S}_{k_S}$ are defined by the following eigenvalue equation\footnote{${k_S}$ is not an index but an eigenvalue.}:
\begin{equation}
\begin{split}
&(\square_{H_{d-1}}+{k_S}^2)\mathbb{S}_{k_S}=0,
\end{split}
\end{equation}
so that it plays a similar role to the one of plane waves $e^{i {k_S} x}$ in planar black holes. 

The EOM \eqref{eqproca} is satisfied if the master variable $\mathcal{A}_{k_S}^L$ satisfies the wave equation,
\begin{equation}
\label{waveeq}
u ^{d-3}D_a \left(\frac{D^a \mathcal{A}_{k_S}^L}{u^{d-3}}  \right)- \frac{{k_S}^2}{u^2}\mathcal{A}_{k_S}^L = 0.
\end{equation}
To obtain the explicit Green's function, let us rewrite above equation more explicitly. By using the Fourier transformation $\mathcal{A}_{k_S}^L(y = \{t,\nr\}) = \int \dd \omega e^{-i \omega t}\mathcal{A}_{k_S}^L(\omega, \nr)$, $u = \cosh \nr$ and the explicit form of $g_{ab}$ \eqref{metricdecom}, \eqref{waveeq} can be written as
\begin{equation}
    \mathcal{A}^{L\prime\prime}(\nr) + \left(\coth{\nr} + (3-d)\tanh{\nr}\right)\mathcal{A}^{L\prime}(\nr)+\left(\frac{\omega^2}{\sinh^2{\nr}}+\frac{L(L+d-2)}{\cosh^2{\nr}}\right)\mathcal{A}^L(\nr)=0,
\end{equation}
where we replace ${A}_{k_S}^L(\omega, \nr) \rightarrow {A}^L(\nr)$ for notational simplicity. The incoming solution to the above equation is
\begin{align} \label{aadd1}
&\mathcal{A}^L\left(z\right) = z^{-\frac{i \omega}{2}}{}_2 F_{1}\left[a, b, a + b +\mathcal{N} + 1, z\right], \nonumber \\
&a = -\frac{1}{2}\left(L+i \omega\right), \quad b = \frac{1}{2}\left(d-2+L-i \omega \right), \quad \mathcal{N} = 1-\frac{d}{2},
\end{align}
where $z=\tanh^2\nr$ and $L$ is defined by ${k_S}^2 =: -L(L+d-2)$\footnote{The change of ${k_S}$ to $L$ is nothing but a convention so that our expression looks more simple.}.

By the same reason explained in the scalar field case, we consider the $\mathcal{N}\in \mathbb{Z}^{*-}$ case and the $\mathcal{N}\notin \mathbb{Z}^{*-}$ case separately. Since $\mathcal{N} = 1 -\frac{d}{2}$ ($\Delta=d-1$), we consider even $d$ and odd $d$ separately.

First, let us consider the case in which $d$ is odd. 
To determine $A_t^{L}$ in \eqref{difgauge} we need to fix the gauge $\mathcal{F}$. We choose the gauge  $\mathcal{F}$ such that $A_{\nr} ^{L}= 0$, which is a usual gauge in the holographic set-up. 
The near boundary behavior of $\mathcal{F}$, which is relevant to the retarded Green's function, is 
\begin{align} \label{FF11}
    \mathcal{F}(z) = &-\frac{i e^{\frac{\pi\omega}{2}}\pi\omega\Gamma(1-i\omega)(1-z)}{2\Gamma(3-\frac{d}{2})\Gamma\left(\frac{1}{2}(-i \omega -L)\right)\Gamma\left(\frac{1}{2}(-i \omega +L+d-2)\right)} +\mathcal{O}(1-z)^2\nonumber \\&+\frac{e^{\frac{\pi \omega}{2}}\pi\omega^2\csc\left(\frac{d \pi}{2}\right)\Gamma(-i \omega)(1-z)^{\frac{d-2}{2}}}{2\Gamma\left(\frac{d}{2}\right)\Gamma\left(\frac{1}{2}(-i\omega-L-d+4)\right)\Gamma\left(\frac{1}{2}(-i\omega+L+2)\right)}+\mathcal{O}(1-z)^{\frac{d-2}{2}+1}\,.
\end{align}
By plugging \eqref{aadd1} and \eqref{FF11} into \eqref{difgauge}, the boundary behavior of $A_t^L$ can be obtained.
The fall off behavior of $A_t^L$ is
\begin{equation}
    A_t^{L} = (1-z)^{0}\left[A(\omega,L) + \cdots \right] +  (1-z)^{\frac{d-2}{2}}\left[B(\omega, L) + \cdots \right].
\end{equation}
From the above result, we obtain the retarded Green's function corresponding to $A_t^{L}$:
\begin{align}
&G^{R}_{\mathrm{longi}}\left(\omega,L\right) \nonumber\\
& \propto \frac{B(\omega, L)}{A(\omega, L)} \propto \frac{\left(L(L+d-2)\right)\Gamma(1-\frac{d}{2})\Gamma(\frac{1}{2}(-i \omega-L))\Gamma\left(\frac{1}{2}(-i \omega+L+d-2) \right)}{ \Gamma(-1+\frac{d}{2})\Gamma\left(\frac{1}{2}(-i \omega-L-(d-4) )\right)\Gamma\left(\frac{1}{2}(-i \omega+L+2)\right)}\,,  \label{jhgf1}
\end{align}
which agrees with $\mathcal{G}_{\mathrm{longi}}^\Delta(\omega_E,k)$ in  \eqref{fouriergauge2pt} after replacing 
$\omega \rightarrow i \omega_E$ and $L \rightarrow \pm i k - \frac{d-2}{2}$ in \eqref{replace123} together with $\Delta \rightarrow d-1$
for the massless case.

This retarded Green's function has the pole-skipping points at
\begin{alignat}{3}
& \omega_* =0 \ \ &&\mathrm{and} \ \  &&L^\pm_* = -\frac{d-2}{2}\mp \frac{d-2}{2}\,,\label{psmll} \\
&\omega_n = -i n  \ \ &&\mathrm{and} \ \ &&L_{n,q} =  -\frac{d-2}{2}\pm \left(-n+2q+\frac{d-6}{2}\right) \,,\label{psmlv}
\end{alignat}
where $n=1,2,\cdots$ and $q =1,2, \cdots ,n$.
The above pole-skipping points are consistent with the field theory results \eqref{pscft} and \eqref{diffsub} for the massless case $\Delta =  d-1$. Here, $L_{n,q}$ is written in a symmetric form for easy comparison with field theory's result. 

Next, for the case  in which $d$ is even, we just present the answer because the procedure is very similar to the case in which $d$ is odd, except that we should use \eqref{hypNI}. The result can be obtained by replacing $\frac{\Gamma(1-d/2)}{\Gamma(-1+d/2)}$ in \eqref{jhgf1} with some digamma functions.\footnote{In principle, there is a possibility that the concrete form of \eqref{jhgf1} will be changed. However, the pole-skipping structure will not be changed.}
See appendix \ref{holoregul} for more details. 
This holographic result is consistent with field theory result \eqref{rvtpf}.


\paragraph{General mass case}
Second, we consider the sector of massive vector perturbations involving the $A_t$ component. As we mentioned at the beginning of this section, the analysis for massive case here is not completely general but it is general enough for our main goal, which is finding the {\it leading} pole-skipping points.

In order to simplify the analysis, we consider perturbations of the form
\be
A_N(t,\nr,\chi,\theta_i) = a_N(t, \nr) G_N( \chi, \theta_i) \,,\label{va}
\ee
where $N=t,\nr,\chi,\theta_i$. 
 With the Fourier transformation $a_N(t,\nr) = \int \dd e^{-i \omega t}a_N(\omega,\nr)$, the equation of motion (\ref{eqproca}) with $N=t$ becomes
\bea
\left[m^2-\frac{L(L+d-2)}{\cosh^2{\nr}} \right] a_t(\nr)+\left[\coth{\nr}+(1-d)\tanh{\nr} \right] a_t'(\nr)-a_t''(\nr)=0,\label{eq-N=t}\,
\eea
where we made the assumption that $G_t(\chi,\theta_i)=Y_{L\,M}^{(d-1)}(i \chi,\theta_i)$ and we replaced $\square_{H_{d-1}}$ with $L(L+d-2)$ by using \eqref{YLM}. In \eqref{eq-N=t}, we set $\omega=0$ for two reasons. First, 
we take advantage of the fact that $\omega =0$ is the first pole-skipping frequency for the vector field. We know that from the field theory analysis of sections \ref{SCTF} and near horizon analysis in \ref{SNH}. Second, if we set $\omega=0$, $a_t$ in the equations of motion is decoupled from the other components of $a_N$ so it alone plays the role of the longitudinal channel. Otherwise, all other fields should be considered together for general mass case analysis.

To solve (\ref{eq-N=t}), we change the variable to $z=\tanh^2\nr$ and use the ansatz 
\be
a_t(z)=(1-z)^{\frac{d-\Delta-1}{2}}F(z)\,, 
\ee
in terms of which the equation of motion becomes
\be
4 (1-z) z F''(z)-2 z (d-2 \Delta +2) F'(z)+ (\Delta +L-1) (d-\Delta +L-1)F(z)=0\,.
\ee
Here we substitute $m$ with $\Delta$ by using $\Delta := (d+\sqrt{(d-2)^2}+4m^2)/2$. A solution to above equation can be expressed in terms of a hypergeometric function as
\begin{align}
\label{masdifhol}
&F(z)={}_{2}F_1\left(a,b,a+b+\delta,z \right)\,,\nonumber \\
&a = \frac{1-L-\Delta}{2}, \quad b = \frac{-1+d+L-\Delta}{2}, \quad \delta = \Delta - \frac{d}{2}\,.
\end{align}
Like the previous sections, if $n$ is zero or positive integer, the boundary behavior of the hypergeometric function is different from the non-integer $\delta$ case. Thus we consider each cases separately. 

First let us focus on $\delta$ is non-integer case. Near the boundary, the above solution can be written as
\be
a_t(z) \approx \left[ A(L)+\cdots \right](1-z)^{\frac{d-\Delta-1}{2}}+\left[ B(L)+\cdots\right](1-z)^{\frac{\Delta-1}{2}}\,,
\ee
where $A$ and $B$ can be calculated by applying \eqref{hypNon} to \eqref{masdifhol}. The zero-frequency Green's function can be obtained as
\be
G^R_{\mathrm{longi}}(L,\omega=0) \propto \frac{B(L)}{A(L)} \propto \frac{\Gamma \left(\frac{d}{2} - \Delta \right) \Gamma \left( \frac{L+\Delta -1}{2}\right) \Gamma \left(\frac{-d-L+\Delta +1}{2} \right)}{\Gamma \left(-\frac{d}{2}+\Delta \right) \Gamma \left(\frac{-L-\Delta +1}{2} \right) \Gamma \left(\frac{d+L-\Delta -1}{2} \right)}\,,
\ee
which is consistent with field theory result \eqref{pscft} up to the replacement
\begin{equation}
    (\omega, L) \rightarrow \left(i \omega_E , \pm ik-\frac{d-2}{2}\right).
    \label{repdif}
\end{equation}
The above replacement can be obtained by comparing the eigenvalues $L(L+d-2)$ and \eqref{eigen123} without $\omega_E$.
Thus, there are the pole-skipping points at
\begin{equation}
\label{mvecpspnonint}
    \omega_* = 0\,, \qquad L^-_*=\Delta-d+1 \,, \ \ \mathrm{or} \ \  L^+_*=1-\Delta \,, 
\end{equation}
which agree with the field theory results \eqref{pscft}.

Next, let us move to the $\delta\in \mathbb{N}$ and $\delta=0$ cases. The computation procedure is similar to the scalar field case \ref{posintescar}. For $\delta\in \mathbb{N}$, $\Delta-d/2$ takes positive integer values so, by using  \eqref{posIhyp}, the retarded Green's function reads
\begin{align}
G^R_{\mathrm{longi}}(L,\omega=0) \propto \frac{B(L)}{A(L)}\propto & \frac{ \Gamma \left( \frac{L+\Delta -1}{2}\right) \Gamma \left(\frac{-d-L+\Delta +1}{2} \right)}{ \Gamma \left(\frac{-L-\Delta +1}{2} \right) \Gamma \left(\frac{d+L-\Delta -1}{2} \right)}\times\nonumber\\
&\left[ \psi\left( \frac{L+\Delta -1}{2} \right)+\psi\left( \frac{-d-L+\Delta +1}{2} \right)\right]\,.
\label{Idifcase}
\end{align}
For $\delta=\Delta - d/2=0$, the retarded Green's function can be calculated by \eqref{zerocase}:
\begin{align}
G^R_{\mathrm{longi}}(L,\omega=0) \propto \psi\left(\frac{1}{2}(1-L-\Delta)\right)+\psi\left(\frac{1}{2}(-1+d-L-\Delta)\right)\,.
\label{Zdifcase}
\end{align}
The above results are consistent with \eqref{rvtpf}  with $\omega_E = 0$ after replacing $L$ with $k$ by using \eqref{repdif}. 


\subsubsection{Transverse channel}\label{retardvectormode} 
\paragraph{General method}
The longitudinal mode in section \ref{retardscalarrmode} can be described by a single master variable~\cite{Ueda:2018xvl, Kodama:2003kk} in the massless case. For the massive case, since the gauge symmetry is broken we should deal with coupled fields. However, for a perturbation which describes the transverse channel, it is possible to write a single equation of motion even in the massive case~\cite{Ueda:2018xvl}.

From \eqref{decomvec}, the general form of the perturbation of the transverse channel can be written as
\begin{equation}
\label{decomvectran}
A_{\mu}dx^{\mu} = A_i^T dx^i, \quad \hat{D}^iA_i^T = 0\,.
\end{equation}
Our perturbation ansatz of the gauge field can be written as
\begin{align}
\label{transgauge}
A_i^T = \sum_{k_V}\mathcal{A}_{k_V}(y) \mathbb{V}_{k_V,i}(x)\,,
\end{align}
where $\mathbb{V}_{k_V,i}$ is an $i$-th component of the vector harmonics which satisfies 
\begin{equation}
\label{vhar}
\left(D_jD^j+k_V^2\right)\mathbb{V}_{k_V,i}=0, \quad D^i \mathbb{V}_{k_V,i} = 0\,.
\end{equation}
Here, $y=(t,\nr)$ and $x$ is the spatial coordinate of the corresponding dual field theory. Compared with the Fourier transformation used in the field theory calculation \eqref{vectorharmon}, the role of $\{{A}_{k_V}(y), \mathbb{V}_{k_V,i}(x)\}$ is similar to the role of $\{v(p),f_i(P;p)\}$ up to a trivial time dependence; $P$ includes time while $x$ does not and $p$ includes $\omega_E$ while $k_V$ does not. 
$k_V^2$ is a real number and, for example, in the specific case of section \ref{transverse2pt} it can be expressed as \eqref{transeigen} without $\omega_E^2$.

Using the above ansatz, the equation of motion for the massive gauge field \eqref{eqproca} becomes
\begin{equation}
\label{vmeq}
{}^{(2)}\square\mathcal{A}_{k_V} + (d-3)\frac{D^a u}{u}D_a \mathcal{A}_{k_V}-\left[\frac{-(d-2) +  k_V^2}{u^2}+m^2\right]\mathcal{A}_{k_V} = 0\,,
\end{equation}
where ${}^{(2)}\square$ denotes the Laplacian operator for $g_{ab}$ and $u = \cosh \nr$. See \eqref{metricdecom} and (\ref{eq-u}).
With the Fourier transformation $\mathcal{A}_{k_V}(y={t,\nr}) = \int \dd \omega e^{-i\omega t}\mathcal{A}_{k_V}(\omega, \nr)$, the above equation becomes
\begin{equation}
\label{vmeqp}
    \mathcal{A}''(\nr)+\left(\coth{\nr}+(d-3)\tanh{\nr}\right)\mathcal{A}'(\nr)+\left(\frac{\omega^2}{\sinh^2\nr}+\frac{(d-2-k_V^2)}{\cosh^2\nr} - m^2\right)\mathcal{A}(\nr)=0\,,
\end{equation}
where we omit the $({k_V},\omega)$ dependence by replacing $\mathcal{A}_{k_V}(\omega, \nr) \rightarrow \mathcal{A}(\nr)$. The solution $\mathcal{A}(\nr)$ reads
\begin{align}
\label{transvecsol}
&\mathcal{A}(z) = (1-z)^{\frac{\Delta-1}2}z^{-\frac{i \omega}{2}}{}_2F_1\left(a, b; a + b + \mathcal{N};z\right)\,, \nonumber \\
&a = \tfrac{1}{2}\left(-i\omega-(L+d-3)+\Delta \right), \quad b = \tfrac{1}{2}\left(-i\omega+L-1+\Delta\right), \quad \mathcal{N} = \frac{d}{2}-\Delta,
\end{align}
where $z = \tanh^2\nr$, $m^2 = (\Delta -1)(\Delta-d+1)$, and $\Delta = (d + \sqrt{(d-2)^2+4m^2})/2 > d/2$. Also we parameterized $k_V^2$ as $k_V^2 = -(L-1)(L+d-3)+1$ because this simplifies the argument of the hypergeometric function and also it is compatible with the expression \eqref{vsh}. {In other words, with $k_V= -(L-1)(L+d-3)+1$, \eqref{vmeqp} and \eqref{eq-G} are the same.}\footnote{In general, for the purpose of the comparison with scalar harmonics the parameterization $k_V^2 = -L(L+d-2)+1$ is more natural~\cite{Ueda:2018xvl}. However, for the purpose of easy comparison with \eqref{vsh} we choose a different parameterization.}

The generic boundary behavior of \eqref{transvecsol} takes following form:
\begin{equation}
\mathcal{A} \approx (1-z)^{\frac{d-\Delta-1}{2}}A(\omega,L)(1-z) + (1-z)^{\frac{\Delta-1}{2}}\Big(B(\omega,L)+C(\omega,L)\log(1-z) \Big)\,.
\end{equation}
Since the asymptotic behavior of all the $A_i$ for non-vanishing $\mathbb{V}_i$ is determined by $\mathcal{A}$,\footnote{See \eqref{transgauge}.} it is enough to know $A$, $B$, and $C$ to calculate each Green's function. Like in section \ref{posintescar}, we consider the $\mathcal{N} \in \mathbb{Z}^{*-}$ case and the $\mathcal{N}\notin \mathbb{Z}^{*-}$ case separately. 

First, for $\mathcal{N}\notin \mathbb{Z}^{*-}$, the retarded Green's function can be read off from \eqref{nIcoef} and \eqref{nIstand}:
\begin{align}
\label{transgeneral}
&G^{R}_{\mathrm{transv}} (\omega, L) \propto \frac{B(\omega, L)}{A(\omega, L)}\nonumber\\ 
& \propto \frac{\Gamma\left(\frac{d}{2}-\Delta\right)\Gamma\left(\frac{1}{2}(-i\omega-(L+d-3)+\Delta)\right)\Gamma\left(\frac{1}{2}(-i\omega+L-1+\Delta)\right)}{\Gamma\left(-\frac{d}{2}+\Delta\right)\Gamma\left(\frac{1}{2}(-i \omega-L+3-\Delta)\right)\Gamma\left(\frac{1}{2}(-i \omega+L+d-1-\Delta)\right)}\,.
\end{align}
To compare this with the field theory result we consider the replacement
\begin{equation} \label{replace12345p}
    (\omega, L) \to \left(i \omega_E, \ \pm i k - {\frac{d-4}{2}}\right) \,,
\end{equation}
which can be seen from the comparison of the eigenvalues \eqref{transeigen} without $\omega_E$ and $-k_V^2 = (L-1)(L+d-3)-1$.
After considering the replacement \eqref{replace12345p}, we can see that the retarded Green's function \eqref{transgeneral} has the same form as \eqref{fouriergauge2ptvector0}, having therefore the same pole-skipping points \eqref{fouriergauge2ptvector0}. In terms of $(\omega, L)$ they are located at
\bea
\label{mvectvpsp}
\omega_n = - i\,n \ \ \mathrm{and}\ \ L_{n,q}=-\frac{d-4}{2}\pm\left(-n+2q+\Delta-\frac{d+2}{2}\right)\,, 
\eea
where $n=1,2,\cdots$ and $q=1,2,\cdots,n$. Here, $L_{n,q}$ are written in a symmetric form for easy comparison with the field theory result. 
The first instance of pole-skipping occurs for $n=q=1$, giving 
\begin{equation}
    \label{mvectvpsplead}
    \omega_* =-i\,, \qquad L^+_*=2-\Delta \,, \ \ \mathrm{or} \ \  L^-_*=\Delta+2-d \,.
\end{equation}

Next, let us move on to the $\mathcal{N}\in\mathbb{Z}^{*-}$ case. For this case, we simply write down the results:
\begin{align} \label{oij8}
G^{R}_{\mathrm{transv}}(\omega, L) \propto & \frac{B(\omega,L)}{A(\omega,L)}\propto \frac{\Gamma \left(\frac{1}{2}(-i\omega-(L+d-3)+\Delta) \right)\Gamma \left(\frac{1}{2}(-i \omega+L-1+\Delta)\right) }{\Gamma \left(\frac{1}{2}(-i \omega-(L-3)-\Delta)\right)\Gamma \left(\frac{1}{2}(-i\omega+L-1-\Delta) \right) } \times \nonumber \\
&\left[\psi\left( \frac{-i\omega-(L+d-3)+\Delta}{2} \right)+\psi\left( \frac{1-i \omega+L-1+\Delta}{2}\right) \right]\,,
\end{align}
for $\mathcal{N}\in \mathbb{Z}^{-}$ \eqref{Istand} and
\begin{equation}
\label{holzerotransvec}
G^{R}_{\mathrm{transv}} (\omega, L)\propto \frac{B(\omega, L)}{A(\omega ,L)} \propto \psi\left(\frac{1}{2}(-i \omega - L - \frac{d-4}{2}+1)\right)+\psi\left(-i \omega + L + \frac{d-4}{2}+1\right),
\end{equation}
for $\mathcal{N} = 0$ \eqref{zerocase}.
The results \eqref{oij8} and \eqref{holzerotransvec} are consistent with the field theory results \eqref{rvtpf} once we take into account the relation \eqref{replace12345p}.

\paragraph{Specific method: $A_{\phi}$ component}
In this section, we derive the retarded Green's function corresponding to excitations of the transverse channel using a very simple ansatz for the vector field.  We first write the hyperbolic space as $dH_{d-1}^2=d\chi^2+\sinh^2{\chi} \, d\Omega_{d-2}^2$, where $d\Omega_{d-2}^2= d\theta^2+\sin^2{\theta}\,d\Omega_{d-3}^2$, with $d\Omega_1=d\phi^2$. We then assume that the only non-zero component of the vector field is $A_{\phi}$, which (for simplicity) does not depend on $\phi$, i.e., $\partial_\phi A_{\phi}=0$. 

The equation of motion for $A_{\phi}$ can then be written as
\be
\label{vmeqtheta}
\partial_{\nr}^2 A_{\phi} -\frac{\partial_t^2 A_{\phi}}{\sinh^2 \nr}+\frac{\square_{H_{d-3}} A_{\phi} }{\cosh^2 \nr}+ \big[ \coth \nr+(d-3) \tanh \nr \big] \partial_{\nr} A_{\phi}-m^2 A_{\phi}=0\,,
\ee
where $\square_{H_{d-3}} := (d-4)\coth{\chi} \partial_{\chi}+\partial_{\chi}^2+\frac{1}{\sinh{\chi}^2}\left[\partial_{\theta}^2+(d-5)\cot{\theta}\, \partial_{\theta}+\cdots \right] $.\footnote{$\square_{H_{d-3}}$ is just a formal definition obtained by replacing $d \to d-2$ in $\square_{H_{d-1}}$.}
Note that the above equation is identical to the equation of motion for the scalar field (\ref{eq-scalar}), with the replacement $d \rightarrow d-2$. Having this in mind, we use the following ansatz
\be
A_{\phi}(t,\nr,\chi,\theta)=\sum_{L,M}G(t,\nr)\, Y_{L\,M}^{(d-3)}(i \chi,\theta) \,. \label{vsh}
\ee
Note that $G(t,\nr)$ and $Y_{L\,M}^{(d-3)}$ are concrete examples of the functions $\mathcal{A}_{k_V}(y)$  and $\mathbb{V}_{k_V,i}(x)$ appearing in \eqref{transgauge}.

By using \eqref{vsh} and $G(t,\nr)=\int \dd \omega e^{-i \omega t}G(\omega, \nr)$ the equation of motion \eqref{vmeqtheta} yields
\be \label{eq-G}
G''(\nr)+\big[ \coth \nr+(d-3) \tanh \nr \big] G'(\nr)+\left[\frac{\omega^2}{\sinh^2\nr}+\frac{L(L+d-4)}{\cosh^2\nr}-m^2 \right]G(\nr)=0\,,
\ee
where we used $\square_{H_{d-3}}Y_{L\,M}^{(d-3)}(i \chi,\theta) = L(L+d-4)Y_{L\,M}^{(d-3)}(i \chi,\theta)$, which can be understood from \eqref{YLM} with $d \to d-2$.  Here, we also replace $G(\omega, \nr)\rightarrow G(\nr)$ for notational simplicity. Because $G$ is nothing but 
$\mathcal{A}$ in \eqref{vmeqp}, all consequences from \eqref{eq-G} are the same as in the previous case.

\section{Near-horizon analysis of the bulk equations of motion\label{SNH}}

In this section, we compute the pole-skipping points for scalar and vectors fields by analyzing the near-horizon bulk equations of motion. Due to the simplicity of the near-horizon equations of motion, this analysis can be done in rather general hyperbolic black holes, as opposed to the exact computation of Green's function performed in Section \ref{SHGF}, which (as far as we know) can only be done for a Rindler-AdS$_{d+1}$ geometry.

Starting from the Einstein-Hilbert action
\begin{equation}
\label{EHaction}
S = \int d^{d+1} x \sqrt{-g}\left[R+\frac{d\left(d-1\right)}{ \ell_{\mathrm{AdS}}^2} \right]\,,
\end{equation}
we consider the following hyperbolic black hole solution
\begin{align}
ds^2&=\frac{\ell_{\mathrm{AdS}}^2}{z^2}\left(-f(z)dt^2+\frac{dz^2}{f(z)} +\ell_{\mathrm{AdS}}^2 \left(d\chi^2+\sinh^2\chi \,d\Omega_{d-2}^2 \right)\right)\,,\label{GHybh}\\
f(z)&=1-\frac{z^2}{\ell_{\mathrm{AdS}}^2}-\left(\frac{z}{z_0}\right)^{d-2} \left(\frac{\ell_{\mathrm{AdS}}^2}{z_0^2}-1\right)\frac{z^2}{\ell_{\mathrm{AdS}}^2}\,,
\end{align}
where $z_0$ denotes the position of the horizon, while the boundary is located at $z=0$. The Hawking temperature is given by
\begin{align}
T=\frac{d-(d-2)z_0^2/\ell_{\mathrm{AdS}}^2}{4\pi z_0 }\,.
\end{align}
By setting $z=\ell_{\mathrm{AdS}}^2/r$ and  $z_0=\ell_{\mathrm{AdS}}$ the metric (\ref{GHybh}) becomes the Rindler-AdS$_{d+1}$ metric (\ref{eq-metric}). From here, we set $ \ell_{\mathrm{AdS}}=1$.
For our purposes, it will be useful to introduce the incoming Eddington-Finkelstein coordinate $v$
\be
v=t+z_*\,,\,\,\,\,\,\,dz_*=-\frac{dz}{f}\,,
\ee
in terms of which the metric becomes
\be
ds^2=-\frac{f(z)}{z^2}dv^2-\frac{2}{z^2} dv dz+\frac{1}{z^2}\left( d\chi^2+\sinh^2\chi d\Omega_{d-2}^2 \right)\,.\label{GHybh2} 
\ee

    In \cite{Blake:2018leo}, the authors found that pole-skipping in energy density two-point functions is related to a special property of Einstein's equations near the black hole's horizon. In general, the Einstein's equations have incoming and outgoing solutions at the horizon. However, at some special value of $(\omega,k)$, one loses a constraint provided by Einstein's equations, and this leads to the existence of two incoming solutions. As a consequence, the corresponding Green's function becomes ill-defined at this special point. It was later observed that pole-skipping also occurs in other sectors of gravitational perturbations, and also for scalar, vector and fermionic fields, being always related to a special property of the near-horizon equations of motion \cite{Blake:2019otz, Grozdanov:2019uhi, Natsuume:2019xcy,Natsuume:2019vcv, Ceplak:2019ymw}. All these studies considered the case of planar black holes, in which the perturbations can be decomposed in terms of plane waves. The study of pole-skipping in hyperbolic black holes was initiated in \cite{Ahn:2019rnq}, in which the authors considered gravitational perturbations related to energy density two-point functions.

Here, we compute the pole-skipping points of two-point functions of scalar and vector fields in the {\it general} hyperbolic black hole metric (\ref{GHybh2}), with a general $z_0 >0$, and show that the leading pole-skipping points in the Rindler-AdS$_{d+1}$ metric, where $z_0 = \ell_\mathrm{AdS} = 1$, agree with the previous field theory results (\ref{eol}) and (\ref{eols}).

\subsection{Scalar field} \label{intro}
We first consider a scalar field with the action (\ref{saction})
propagating in the background (\ref{GHybh2}).
In terms of the coordinates $(v,z,x_i)$, where $x_i \in \mathbb{H}^{d-1}$, the equation of motion for the massive scalar field, $\left(\square - m^2 \right)\phi(x) = 0$, can be written as
\begin{equation}
\label{KGeq}
z^{d+1} \partial_z \left(z^{1-d} f(z) \partial_z \phi \right)+z(d-1) \partial_{v} \phi
-2z^2 \partial_v \partial_z \phi+z^2 \square_{H_{d-1}}\phi -m^2 \phi=0\,. 
\end{equation}
The above equation can be solved by decomposing the perturbation in terms of hyperspherical harmonics
\be
\phi(v,z,x^i)= \sum_{L,M}\Phi(t, z) Y_{L\,M}^{(d-1)}(i \chi, \Omega_i)\,,
\ee
where $\Omega_i \in S^{d-2}$. With the above ansatz and $\Phi(t,z) = \int \dd \omega e^{-i \omega v}\Phi(\omega, z)$, the equation of motion boils down to
\be \label{eqn1235}
z^{d+1} \partial_z \left(z^{1-d} f(z) \partial_z \Phi \right)-z(d-1) i \omega \Phi
+2z^2 i \omega \partial_z \Phi+z^2 L(L+d-2)\Phi -m^2 \Phi=0\,, 
\ee
where we use $\square_{H_{d-1}}Y_{L\,M}^{(d-1)}(i \chi, \Omega_i)=L(L+d-2)Y_{L\,M}^{(d-1)}(i \chi, \Omega_i)$. With the expansion
\begin{equation} \label{expan123}
  \Phi(z)=\sum_{j}^{\infty} \Phi_j (z-z_0)^j \,, 
\end{equation}
in the near-horizon limit, the leading order terms, $(z-z_0)^0$, of the equation \eqref{eqn1235} becomes
\begin{equation}
\label{GNHS}
\left(2i \omega-4\pi T \right)\Phi_1-\left(\frac{i\left(d-1\right)\omega}{z_0}+\frac{\Delta\left(\Delta - d\right)}{z_0^2}-L\left(L+d-2\right)\right)\Phi_0 = 0\,,
\end{equation}
where $m^2 = \Delta(\Delta-d)$.
For generic values of $\omega$ and $L$ except
\begin{equation} \label{spepnt}
\begin{split}
\omega_*&=-i 2\pi T\,, \\
L_*^{\pm}&=-\frac{(d-2) z_0 \pm \sqrt{d^2 \left(2-z_0^2\right)+2 d \left(z_0^2-2 \Delta -1\right)+4 \Delta ^2}}{2 z_0}.
\end{split}
\end{equation}
\eqref{GNHS} provides a constraint between $\Phi_1$ and $\Phi_0$. In other words, (\ref{GNHS}) fixes $\Phi_1$ in terms of $\Phi_0$. Furthermore, by solving the equation of motion at order $(z-z_0)^j$ we can fix $\Phi_{j+1}$ in terms of $\Phi_{j}$. This implies that all higher-order terms $\Phi_j$ can be fixed in terms of $\Phi_0$, and we can find a regular solution for $\Phi(z)$ that is unique up to a overall normalization. 

However, at special points $(\omega_*,L_*)$ we lose the constraint between $\Phi_0$ and $\Phi_1$ and both are free and arbitrary.
 This leads to the existence of two regular solutions which are consistent with incoming boundary conditions at $(\omega,L)=(\omega_*,L_*)$. Two arbitrary free parameters yield an ``ambiguity or non-uniqueness'' in the corresponding
retarded Green's function.\footnote{For arbitrary values of $\omega$ and $L$, we also have a second solution to the equations of motion, but this solution is not regular at the horizon, corresponding to the outgoing solution.  At $(\omega,L)=(\omega_*,L_*)$, this second solution becomes regular.}.
Intuitively, one possible way to have a ``non-unique'' value is $ \frac{0}{0}$, which is nothing but what we have done in the pole-skipping analysis. Thus, we may understand why there is a relation between the special point such as \eqref{spepnt} and the pole-skipping point.

At $z_0=1$, the geometry reduces to a Rindler-AdS$_{d+1}$ geometry, and the first pole-skipping point becomes
\bea \label{eq-nha-scalar}
\omega_*&=&-i \,, \nonumber\\
L_*^{+}&=&1-\Delta\,, \quad L_*^{-}=\Delta-d+1\,,
\eea
which agrees with the results \eqref{scalarpspcft} and \eqref{scalarpsp} obtained in the previous sections.  In this paper, we compute only the leading pole-skipping point by the near horizon analysis. However, it is also possible to obtain other sub-leading pole-skipping points by considering the near-horizon equation of motion \eqref{eqn1235} with higher orders terms in the near-horizon expansion.
For example, if we consider the equation \eqref{eqn1235} up to $(z-z_0)$ order, three coefficients $\Phi_0, \Phi_1, \Phi_2$ in \eqref{expan123} are involved. In this case, the condition that $\Phi_0, \Phi_2$ are free gives the second pole-skipping points.
We refer to \cite{Blake:2019otz} for more details.

\subsection{Vector field} 
In this section, we consider the vector field action (\ref{paction}) which propagates in the background (\ref{GHybh2}), where $F_{\mu\nu} = \partial_{\mu} A_{\nu} - \partial_{\nu} A_{\mu}$, and $m$ is the mass of the vector field.
The equations of motion resulting from the above action are (\ref{eqproca}). In order to simplify the analysis, we consider perturbations that do not depend on the coordinates $\theta_i$ on $S^{d-2}$, i.e., $A_M = A_M (v,z,\chi)$.
In this case, the equations of motion for $A_{\theta_i}$ (transverse channel) decouple from the other equations of motion, and the components $A_v, A_z$, and $A_{\chi}$ form an independent sector (longitudinal channel). Let us start with the longitudinal channel.

\subsubsection{Longitudinal channel}\label{horizonscalarmode} 
The equations of motion for $A_v, A_z$, and $A_{\chi}$ in the longitudinal channel read:
\begin{equation} \label{ordfg}
\begin{split}
&\frac{m^2}{z} A_z -z \, \square_{H_{d-1}} A_z-z \big[(d-2) \coth{\chi}\, \partial_z A_{\chi}+\partial_{\chi} \partial_{z} A_{\chi} \big]+z \partial_z^2 A_v \\
& \qquad\qquad +(d-3)(\, \partial_v A_z-\partial_z A_v)-z \partial_v \partial_z A_z=0\,, \\
&\frac{m^2}{z^2} \left(A_v -f A_z \right)- f\left[ (d-2) \coth{\chi}\, \partial_z A_{\chi}+\partial_{\chi} \partial_{z} A_{\chi} \right]-\square_{H_{d-1}} A_v+ f\, \square_{H_{d-1}} A_z \\ 
& \qquad\qquad +\left[ (d-2) \coth{\chi}\, \partial_v A_{\chi}+\partial_{\chi} \partial_{v} A_{\chi} \right]+\partial_v \partial_{z} A_{v}-\partial_v^2 A_z=0\,, \\
& -\frac{m^2}{z} A_{\chi}+f\, \left[(d-3)(\partial_{\chi}A_z)-\partial_z A_{\chi}+z (\partial_z^2 A_{\chi}-\partial_z \partial_{\chi}A_z) \right] \\
& \qquad\qquad +(d-3)\left(\partial_{v}A_{\chi} -\partial_{\chi}A_v\right) 
+ z\left[f'(\partial_z A_{\chi}-\partial_{\chi}A_z) \right. \\ 
& \qquad\qquad  + \left.\partial_z \partial_{\chi}A_v+ \partial_v \partial_{\chi}A_z-2 \partial_v \partial_{z}A_{\chi}\right]=0\,.\\
\end{split}
\end{equation}
After a careful inspection of the above equations of motion, we choose the following ansatz
\begin{equation} \label{ans123}
    \begin{split}
A_v&= \sum_{L}a_v(v,z) \, Y_{L\,0}^{(d-1)}(i \chi,0)\,,\\
A_z&= \sum_{L} a_z(v,z)\,  Y_{L\,0}^{(d-1)}(i \chi,0)\,,\\
A_{\chi}&= \sum_{L} a_{\chi}(v, z) \, \partial_{\chi} Y_{L\,0}^{(d-1)}(i \chi,0)\,.
    \end{split}
\end{equation}
Here, we chose $A_{\chi} \propto \partial_{\chi} Y_{L\,0}^{(d-1)}(i \chi,0)$\footnote{Here $Y_{L\,0}^{(d-1)}(i \chi,0)$ means $(d-1)$ dimensional spherical harmonics which depends on only $\chi$.  Note that the subscript $0$ in $Y_{L0}$ is important. For example, recall that for the usual 3-dimensional spherical harmonics: $Y_{LM}(\theta,0)=P_L^M(\cos\theta)\neq P_L^0(\cos \theta)=Y_{L0}(\theta,0)$.} to make the operator $(d-2)\coth{\chi} \partial_{\chi} +\partial_{\chi}^2$ appear acting on $\partial_z A_{\chi}$ in the first equation and $\partial_v A_{\chi}$ in the second equation. In angular independent perturbations, the above operator equals $\square_{H_{d-1}}$, which has $Y_{L\,0}^{(d-1)}(i \chi,0)$ as an eigenfunction with eigenvalue $L(L+d-2)$. With the ansatz \eqref{ans123} and the Fourier transformations $a_N(v,z) = \int \dd \omega e^{-i\omega v} a_N(\omega, z)$ where $N = v,z, \chi$, the equations \eqref{ordfg} boil down to
\begin{equation}
\begin{split}
&\left[m^2-z^2 L(L+d-2)+i (d-3)\omega \right]a_z \\
&\qquad\qquad +z \Big[ (d-3)a_v'-i \omega a_z'
+L(L+d-2)a_{\chi}' -a_v''
\Big] =0\,,\\
&\left[m^2-z^2 L(L+d-2) \right] a_v+\big[ z^2 \omega^2+f\big(L(L+d-2)-m^2)\big]a_z \\
& \qquad\qquad -iz^2 \big[ \omega a_v'+L(L+d-2)(\omega a_{\chi}-i f a_{\chi}') \big]=0\,, \\
&\left[-m^2-i (d-3)z \omega \right] a_{\chi}+z \big[-(d-3)a_v+a_z \left[(d-3)f -z f' -i \omega z \right] \\
& \qquad\qquad +z a_v'-z f a_z'+\left[-(d-3)f+z(2 i \omega+f') \right]a_{\chi}'+z f a_{\chi}''\big]=0\,.
\end{split}
\end{equation}

By considering the following near-horizon expansion
\begin{equation}
    \begin{split}
&a_v(z)=a_v^{(0)}+a_v^{(1)}(z-z_0)+a_v^{(2)}(z-z_0)^2 +\cdots,  \\
&a_z(z)=a_z^{(0)}+a_z^{(1)}(z-z_0)+a_z^{(2)}(z-z_0)^2 + \cdots,  \\
&a_{\chi}(z)=a_{\chi}^{(0)}+a_{\chi}^{(1)}(z-z_0)+a_{\chi}^{(2)}(z-z_0)^2  +\cdots,
\end{split}
\end{equation}
the equations of motion become 
\begin{equation} \label{mnbg}
\begin{split}
& a_z^{(0)} \left[m^2-L(L+d-2)z_0^2+i(d-3)\omega \right] \\
& \qquad\qquad +z_0 \left[(d-3)a_v^{(1)}+z_0 \left(a_{\chi}^{(1)}L(L+d-2)-2a_v^{(2)}-i a_{z}^{(1)} \omega \right)  \right]=0\,,  \\
& a_v^{(0)}\left[m^2-L(L+d-2)z_0^2 \right]+z_0^2 \omega \left[\omega a_z^{(0)}-i \left(a_v^{(1)}+a_{\chi}^{(0)} L(L+d-2) \right) \right]=0\,, \\
& a_{\chi}^{(0)}m^2-z_0 \left(a_v^{(1)}z_0-(d-3)a_v^{(0)} \right) \\
& \qquad\qquad +i z_0 \left[(d-3)a_{\chi}^{(0)}+(a_z^{(0)}-2a_{\chi}^{(1)})\omega +(a_{\chi}^{(1)}-a_{z}^{(0)})z_0^2 f'(z_0) \right]=0\,.
\end{split}
\end{equation}
\eqref{mnbg} together with the Lorentz condition give the relation between the three leading coefficients ($a_i^{(0)}$) and  the four sub-leading coefficients ($a_i^{(1)},a_v^{(2)}$) in general. In other words, once we fix ($a_i^{(0)}$), ($a_i^{(1)}, a_v^{(2)}$) are determined. 
However, for the following particular values of $\omega$ and $L$

\be\label{GP}
\omega =0\,, \qquad L=L_*^{\pm} :=- \frac{1}{2} \left(d-2 \pm \sqrt{(d-2)^{2} + \frac{4m^2}{z_{0}^2}} \right) \,,
\ee
the second equation is trivially satisfied and we lose one constraint. This means that there is no specific relation between $a_v^{(0)}$ and $a_v^{(1)}$ similarly to the absence of relation between $\Phi_0$ and $\Phi_1$ that takes place at the pole skipping point in the case of the scalar field.

Finally, by setting $z_{0}=1$ into \eqref{GP}, we recover the Rindler-AdS result, namely
\be
L_*^{+} \,=\,   1 - \Delta    \,, \qquad L_*^{-} \,=\,  1-d+\Delta  \,,\label{neardiffu}
\ee
where we wrote the mass in terms of the scaling dimension $m^2 = (\Delta-1)(\Delta-d+1)$.
 The above results agree with the results \eqref{mvecpspnonint} and \eqref{pscft} obtained in the previous sections.

\subsubsection{Transverse channel}\label{horizonvectormode} 
The equation of motion for $A_{\theta}$ greatly simplifies in the case of perturbations that do not depend on the coordinates on $S^{d-2}$. In the coordinates $(v,z,x_i)$ defined in (\ref{GHybh2}), the equation of motion for $A_{\theta}$ becomes
\bea
z^2 f(z) \partial_z^2 A_{\theta}+z\left[z \partial_z f(z)-(d-3)f(z) \right]\partial_z A_{\theta}
-(d-3)\partial_{v} A_{\theta}-2z^2 \partial_v \partial_z A_{\theta}\nonumber \\+z^2\left[(d-4)\coth{\chi} \partial_{\chi}+\partial_{\chi}^2 \right] A_{\theta}-m^2 A_{\theta}=0\,.
\eea
Since $\left[(d-4)\coth{\chi} \partial_{\chi}+\partial_{\chi}^2 \right] A_{\theta} = \square_{H_{d-3}} A_{\theta}$, we use the following ansatz
\be
A_{\theta}= \sum_L \, a_{\theta}(v,z) \, Y_{L\,0}^{(d-3)}(i \chi, 0)\,,
\ee
where $a_{\theta}(v,z)$ is a function to be determined. With the above ansatz and $a_{\theta}(v,z) = \int \dd \omega e^{-i\omega v}a_\theta(\omega,z)$, the equation of motion becomes
\be
z^2 f a_{\theta}''+z^2\left[(2 i \omega +f')-(d-3)\frac{f}{z} \right]a_{\theta}'+\left[ z^2 L(L+d-4)-i (d-3)\omega -m^2 \right]a_{\theta}=0\,,
\ee
where the primes denote derivatives with respect to $z$. 
In the near-horizon limit, $z\rightarrow z_0$, the above equation becomes
\be
z_0^2\left[(2 i \omega +4 \pi T)\right]a_{\theta}'(z_0)+\left[ z_0^2 L(L+d-4)-i (d-3)\omega -m^2 \right]a_{\theta}(z_0)=0\,,
\ee
where we used that $f(z_0)=0$ and $f'(z_0)=4 \pi T$.

Again, we can identify the leading pole-skipping point as the value of $(\omega,L)$ such that the coefficients of both $a_{\theta}(z_0)$ and $a_{\theta}'(z_0)$ are zero. This happens for
\bea
\omega_* &=& - i 2 \pi T\,,\nonumber\\
L_*^{\pm}&=&-\frac{(d-4) z_0 \pm \sqrt{d^2 \left(2-z_0^2\right)+2 d \left(-2 \Delta +z_0^2-1\right)+4 \left(\Delta ^2+z_0^2-1\right)}}{2 z_0}\,,
\eea
where we use that $m^2=(\Delta-1)(\Delta+1-d)$.
The special case of a Rindler-AdS geometry is obtained by setting $z_0=1$
\bea
\omega_* &=& - i\,,\nonumber\\
L_*^{+}&=&2-\Delta\, \quad L_*^{-}=2-d+\Delta\,.\label{neartransv}
\eea
The above result perfectly matches the results obtained in the previous sections \eqref{mvectvpsp}, \eqref{slvpsp}.

\section{Conclusions\label{discussion}}
We have studied the pole-skipping points in the momentum space ($\omega, L$) of two point Green's functions in hyperbolic space. One intuitive way to describe the pole-skipping phenomena is as follows. In the momentum space ($\omega, L$), there are continuous lines yielding poles of the Green's functions, except at some discrete points. {These} discrete points {are  called} ``pole-skipping'' points. Furthermore, at these points, the Green's function is not uniquely defined.

Inspired by the analysis of the energy momentum tensor operator in hyperbolic space \cite{Haehl:2019eae}, we have investigated the cases with scalar and vector operators. 
One of the motivations to study the hyperbolic space is that in this case we can use some analytic formulas in our field theory and holographic calculations, which allows us to compare both results analytically.   
We computed the pole-skipping points by three methods: i) conformal field theory analysis, ii) {exact calculation of holographic Green's functions}, and iii) near horizon analysis of the dual geometry. We confirmed explicitly that all methods give the same results. 

Furthermore, we have shown, via conformal block analysis, that the ``leading'' pole-skipping points (${\omega}_*, {L}_*$), meaning the one with the largest imaginary value of $\omega$ among all pole-skipping points\footnote{Since Im(${\omega}_*$)$\leq 0$ for all pole-skipping points, the leading pole skipping points in scalar and vector fields are the ones  with the smallest absolute value of $\omega_*$.}, can be captured by the late time ($t \gg 1$) and large distance ($\mathbf{d} \gg 1$) limit of the OTOC four point function 
\begin{align}
& \langle V(t, \mathbf{d})W(0,0)V(t, \mathbf{d})W(0,0)\rangle_{\ell,\Delta} \sim e^{-i{\omega}_* t+ {L}_*\mathbf{d}}\,, \\
&  \ \ {\omega}_* = i (\ell-1)\,, \quad {L}_*^+= 1-\Delta  \  \ \mathrm{and} \ \ {L}_*^-=\Delta -d + 1 \,,
\label{dolxx}
\end{align}
where $\ell$ and $\Delta$ denote the spin and conformal dimension of the exchange operator of the given conformal block. 
In the exponents, (${\omega}_*, {L}_*$) are nothing but the leading pole-skipping points and analogues of the ``Lyapunov exponent'' and ``butterfly velocity'' obtained in the case where the exchange operator is the energy momentum tensor.  

Note that in section \ref{section3} we do not need the assumption of holographic CFT\footnote{We need to assume holographic CFT to choose a dominant conformal block with energy momentum tensor exchange.} to show the relation between conformal blocks and pole-skipping points of conformal two point functions because they are universal objects in any CFT. Holographic methods in section \ref{SHGF} and \ref{SNH} provide an alternative and nontrivial check.

Here, we summarize our results for the pole-skipping points ($d\geq3$). 
\subsubsection*{Scalar field}
The pole-skipping points of a scalar field in both the field theory and holography are \eqref{scalarpsp}:
\bea
\omega_n = - i\,n \ \ \mathrm{and}\ \ L_{n,q}=-\frac{d-2}{2}\pm\left(-n+2q+\Delta-\frac{d+2}{2}\right)\,,
\eea
where $n=1,2,\cdots$ and $q=1,2,\cdots,n$ with the unitarity bound $\Delta>\frac d2-1$ ($\Delta=\frac d2-1$ is excluded by the argument below \eqref{plhyt}).
We obtain the field theory results \eqref{sps} by replacing $\omega \rightarrow i \omega_E$ and $L \rightarrow \pm i k - \frac{d-2}{2}$.

The leading pole skipping points occur for\footnote{Recall that $2\pi T =1$.} 
\begin{equation} \label{con1230}
    \omega_* =-i\,, \qquad  L_*^+=1-\Delta \, \ \ \mathrm{or} \ \  L_*^-=\Delta-d+1\,.
\end{equation}
The same leading pole skipping points have  also been obtained by a near-horizon analysis (see (\ref{eq-nha-scalar})).

\subsubsection*{Vector field: longitudinal channel}  In the case of the longitudinal mode of the vector field, we find the pole-skipping points for general $\Delta$ and $d$ by field theory calculations (\ref{pscft}, \ref{diffsub}):
\begin{alignat}{3}
    & \omega_{E*} =0\ \ &&\mathrm{and} \ \  &&ik_* = \pm\left(\Delta-\frac{d}{2}\right)\,, \label{con12340} \\
    &\omega_{E,n}=-n\ \ &&\mathrm{and}\ \ 
     &&ik_{n,q}=\pm\left(-n+2q+\Delta-\frac{d+4}{2}\right)\,,
\end{alignat}
where $n=1,2,\cdots$ and $q=1,2,\cdots,n$ with the unitarity bound $\Delta\geq d-1$. On the other hand, in the holographic calculation, due to technical complications, we compute the pole-skipping points considering two cases: i) massless vector fields; ii) leading pole skipping point for massive vector fields.

First, for a {\it massless} ($\Delta = d-1$) vector field (\ref{psmll},\ref{psmlv}) we obtain:
\begin{alignat}{3}
& \omega_* =0 \ \ &&\mathrm{and} \ \  &&L^\pm_* = -\frac{d-2}{2}\mp \frac{d-2}{2}\,, \\
&\omega_n = -i n  \ \ &&\mathrm{and} \ \ &&L_{n,q} =  -\frac{d-2}{2}\pm \left(-n+2q+\frac{d-6}{2}\right) \,,
\end{alignat}
where $n=1,2,\cdots$ and $q =1,2, \cdots ,n$ which agree with the field theory's calculation if we replace $\omega \rightarrow i \omega_E, L \rightarrow \pm i k - \frac{d-2}{2}$ and impose the massless vector condition $\Delta=d-1$.

Next, for a {\it massive} vector field, we only compute the leading pole-skipping points, which agree with the field theory result. It turns out that the leading pole-skipping points are \eqref{mvecpspnonint}:
\begin{equation}
    \omega_*= 0\,, \qquad  L^+_*=1-\Delta\,, \ \ \mathrm{or} \ \  L^-_*=\Delta-d+1 \,.
\end{equation}
The same leading pole-skipping points have  been obtained also by a near-horizon analysis (see (\ref{neardiffu})).

\subsubsection*{Vector field: transverse channel}
In the case of the transverse mode of the vector field, we find the pole-skipping points \eqref{mvectvpsp}:
\bea
\omega_n = - i\,n \ \ \mathrm{and}\ \ L_{n,q}=-\frac{d-4}{2}\pm\left(-n+2q+\Delta-\frac{d+2}{2}\right)\,,
\eea
where $n=1,2,\cdots$ and $q=1,2,\cdots,n$ with the unitarity bound $\Delta\geq d-1$.
We can obtain the field theory formula \eqref{scalarpspcft} by replacing $\omega \rightarrow i \omega_E, L \rightarrow \pm i k - \frac{d-4}{2}$.

The leading pole skipping points occur at \eqref{mvectvpsplead}:
\begin{equation}
    \omega_* =-i\,, \qquad    L^+_*=2-\Delta\,, \ \ \mathrm{or} \ \ L^-_*=\Delta-d+2\,.
\end{equation}
The same leading pole skipping points have  also been obtained by a near-horizon analysis (see (\ref{neartransv})).
Even though this is the leading pole-skipping point in the transervse channel of the vector field, it is not the leading of the vector field because the longitudinal channel has a leading pole at $\omega_* = 0$. 

\subsubsection*{Energy density $(\ell=2, \Delta=d)$}
For the convenience of the readers, we also summarize the leading pole-skipping points in energy density two point function from previous works. The leading pole-skipping points in hyperbolic space are
\begin{equation} 
    \omega_* =+i\,, \qquad  L_*^+=1-d \, \ \ \mathrm{or} \ \  L_*^-=1\,.
\end{equation}
These points were derived from a field theory calculation \cite{Haehl:2019eae} and from a near-horizon analysis for the sound channel \cite{Ahn:2019rnq}. This result agrees with (\ref{dolxx}).

We have shown that the leading pole-skipping points are linked to the behavior of conformal blocks and their shadow conformal block with an analytic continuation for OTOCs. Because a conformal four point function,  after inserting the projector with two point function, is related with a linear combination of conformal block and its shadow conformal block \eqref{IFP} it is natural to expect that the shadow conformal block is also related with pole-skipping points like the conformal block.

Let us discuss some future directions regarding this work. Based on the relation between conformal blocks and pole-skipping points for scalar and vector fields, which we have shown, and for the energy momentum tensor \cite{Haehl:2019eae}, one can expect a similar relation to hold for symmetric traceless tensor fields with arbitrary spin $\ell$. Computations of the pole-skipping points for general $\ell$ would be straightforward but tedious. It would be useful to develop an easy way of computing pole-skipping points. In our computations, it is unclear why the pole-skipping points are related to the behavior of conformal blocks. Integral representations of conformal blocks might be useful to understand the origin of this relation.

It would be interesting to generalize our work to other fields such as fermion and anti-symmetric tensor fields. For example, the pole-skipping points of fermion fields were studied by holography in \cite{Ceplak:2019ymw}. 
It is also interesting to investigate the pole-skipping phenomena in QFTs without conformal symmetry. Computations of the late time behavior of basis for correlation functions in QFTs, which corresponds to the conformal block in CFTs, might be important for studying the pole-skipping phenomena in general QFTs.

In a holographic computation for $\mathcal{N}=4$ SYM in flat space, it was seen that the leading pole-skipping points in all three channels of energy momentum tensor correlators occur with the same absolute values \cite{Grozdanov:2019uhi}. In our results for vector fields, the leading pole-skipping points of the two channels in hyperbolic space are not correlated. However, one can observe a coincidence between the sub-leading pole-skipping points. In particular,  the sub-leading pole-skipping points (6.6) and (6.10) with lower minus sign are identical. For the energy momentum tensor,  gravitational perturbations that correspond to the sound channel in hyperbolic black holes have been analyzed in \cite{Ahn:2019rnq}. It would be interesting to investigate if the pole-skipping points of other sectors of gravitational perturbations are also correlated to the pole-skipping points of the sound channel.

Another interesting future direction would be to investigate if it is possible to have some form of diffusive or hydrodynamic modes in hyperbolic space, and how this is related to pole-skipping. We hope to address this question in the near future.

\acknowledgments
We would like to thank Felix M. Haehl, Ioannis Papadimitriou and Chang-Woo Ji for valuable discussions and comments. We are indebted to an anonymous referee
for helpful suggestions and comments.
This work was supported by Basic Science Research Program through the National Research Foundation of Korea(NRF) funded by the Ministry of Science, ICT \& Future
Planning (NRF-2017R1A2B4004810) and the GIST Research Institute(GRI) grant funded by the GIST in 2020. We also would like to
thank the APCTP(Asia-Pacific Center for Theoretical Physics) focus program, ``Quantum Matter from the Entanglement and Holography" in Pohang, Korea for the hospitality during our visit, where part of this work was done.

\appendix 
\section{Monodromy of the Hypergeometric function}\label{mhgf}
We start from a formula of the Hypergeometric function \cite{silverman1972special}
\begin{align}
_2F_1(\alpha, \alpha; 2\alpha; 1-v)=\frac{\Gamma(2\alpha)}{\Gamma(\alpha)^2}\sum_{k=0}^\infty\frac{(\alpha)_k(\alpha)_k}{(1)_k}[2\psi(k+1)-2\psi(\alpha+k)-\log v ]\frac{v^k}{k!}\,,
\end{align}
where $(\alpha)_k:=\Gamma(\alpha+k)/\Gamma(\alpha)$ is the Pochhammer symbol, and $\psi(k):=\frac{\frac{d}{dk}\Gamma(k)}{\Gamma(k)}$ is the digamma function.  To avoid the divergence of  $\Gamma(\alpha)$, we assume $\alpha\ne 0, -1, -2,\cdots$. If $\alpha$ is a nonpositive integer, the hypergeometric function reduces to a polynomial, and we cannot pick up the monodromy along $v\to e^{-2\pi i}v$. Because of multi-valuedness of $\log v$, we obtain
\begin{align}
&_2F_1(\alpha, \alpha; 2\alpha; 1-e^{-2\pi i}v)-\;_2F_1(\alpha, \alpha; 2\alpha; 1-v)\notag\\
=&2\pi i\frac{\Gamma(2\alpha)}{\Gamma(\alpha)^2}\;_2F_1(\alpha, \alpha; 1; v)\,.\label{hgf1}
\end{align}
To obtain the behavior around $v=1$, we use \cite{silverman1972special}
\begin{align}
_2F_1(\alpha, \alpha; 1; v)\notag
=&\frac{\Gamma(1-2\alpha)}{\Gamma(1-\alpha)^2}\;_2F_1(\alpha, \alpha; 2\alpha; 1-v)\notag\\
+&\frac{\Gamma(2\alpha-1)}{\Gamma(\alpha)^2}(1-v)^{1-2\alpha}\;_2F_1(1-\alpha, 1-\alpha; 2(1-\alpha); 1-v)\,,\label{hgf2}
\end{align}
where we assume that $2\alpha-1$ is not an integer. By using (\ref{hgf1}) and (\ref{hgf2}), we can determine the constants $A$ and $B$ in (\ref{hgf3}) as
\begin{align}
A=&2\pi i\frac{\Gamma(\Delta+\ell)\Gamma(1-(\Delta+\ell))}{\Gamma((\Delta+\ell)/2)^2\Gamma(1-(\Delta+\ell)/2)^2},\notag\\
B=&2\pi i\frac{\Gamma(\Delta+\ell)\Gamma(\Delta+\ell-1)}{\Gamma((\Delta+\ell)/2)^4}\,.
\end{align}

\section{Fourier transformation of scalar two point function in hyperbolic space }\label{integration}

The Fourier transformation of the parameterized scalar two point function \eqref{ptp} has been done in \cite{Haehl:2019eae,Ohya:2016gto}. Here, we reproduce the computation for the reader's convenience. Starting from the formal expression
\be
    \mathcal G_{(a,b)}^\Delta f(P;\omega_E,k,\vec p_\perp)=\int dP' \mathcal G_{(a,b)}^\Delta(P,P')f(P';\omega_E,k,\vec p_\perp)\,,
\ee
we can explicitly integrate above by plugging the eigenfunction $f(P;\omega_E,k,\vec p_\perp)$ \eqref{eigen1}

\begin{align}
\mathcal G_{(a,b)}^\Delta&(\omega_E,k)\rho^{\frac{d-2}{2}}K_{ik}(|\vec p_\perp|\rho)\nonumber\\
&=\int dP' \mathcal G_{(a,b)}^\Delta(P,P')\rho'^{\frac{d-2}{2}}K_{ik}(|\vec p_\perp|\rho')e^{i\omega_E(\tau'-\tau)+i\vec p_\perp\cdot (\vec x_\perp'-\vec x_\perp)}\nonumber\\
&=\int_0^\infty\frac{d\rho'}{\rho'^{d-1}}\int_0^{2\pi}d\tau'\int d^{d-2}\vec x'_\perp\frac{1}{\left(-2a\cos(\tau-\tau')+b\frac{\rho^2+\rho'^2+(\vec x_{\perp}-\vec x'_{\perp})^2}{\rho\rho'}\right)^{\Delta}}\nonumber\\
&\qquad\qquad\qquad\times\rho'^{\frac{d-2}{2}}K_{ik}(|\vec p_\perp|\rho')e^{i\omega_E(\tau'-\tau)+i\vec p_\perp\cdot (\vec x_\perp'-\vec x_\perp)}\nonumber\\
&=\frac{1}{2^\Delta\Gamma(\Delta)}\int_0^\infty d\zeta\,\zeta^{\Delta-1}\int_0^\infty\frac{d\rho'}{\rho'^{d-1}}\left[\int d^{d-2}\vec x'_\perp e^{-b\frac{(\vec x_{\perp}-\vec x'_{\perp})^2}{2\rho\rho'}\zeta+i\vec p_\perp\cdot (\vec x_\perp'-\vec x_\perp)}\right]\nonumber\\
&\qquad\qquad\qquad\times\left[\int_0^{2\pi}d\tau'e^{i\omega_E(\tau'-\tau)}e^{a\cos(\tau-\tau')\zeta}\right]e^{-b\frac{\rho^2+\rho'^2}{2\rho\rho'}\zeta}\rho'^{\frac{d-2}{2}}K_{ik}(|\vec p_\perp|\rho')\nonumber\\
&=\frac{(2\pi)^{\frac{d}{2}}\rho^{\frac{d-2}{2}}}{2^\Delta\Gamma(\Delta)} b^{-\frac{d-2}2}\int_0^\infty d\zeta\,\zeta^{\Delta-\frac{d}{2}}\,I_{\omega_E}(a\zeta)\left[\int_0^\infty \frac{d\rho'}{\rho'}\,e^{-b\frac{\rho^2+\rho'^2}{2\rho\rho'}\zeta-\frac{\rho\rho'}{2b\zeta}|\vec p_\perp|^2}K_{ik}(|\vec p_\perp|\rho')\right]\nonumber\\
&=\frac{(2\pi)^{\frac{d}{2}}\rho^{\frac{d-2}{2}}K_{ik}(|\vec p_\perp|\rho)}{2^{\Delta-1}\Gamma(\Delta)}b^{-\frac{d-2}2}\left[\int_0^\infty d\zeta\,\zeta^{\Delta-\frac{d}{2}}\,I_{\omega_E}(a\zeta)K_{ik}(b\zeta)\right]\nonumber\\
&=\frac{\pi^{\frac{d}{2}}}{\Gamma(\Delta)}\frac{a^{\omega_E}}{b^{\Delta+\omega_E}}\frac{\rho^{\frac{d-2}{2}}K_{ik}(|\vec p_\perp|\rho)|\Gamma(\alpha)|^2}{\Gamma(\alpha+\alpha^*+\frac{d}{2}-\Delta)}\;_2F_1\left(\alpha,\alpha^*;\alpha+\alpha^*+\frac{d}{2}-\Delta;\frac{a^2}{b^2}\right)\,,\label{longcomp}
\end{align}
where $\alpha:= \frac{1}{2}(\omega_E+ik-\frac{d-2}{2}+\Delta)$ (see the details in \cite{Haehl:2019eae, Ohya:2016gto}). In the third equality, we used the Schwinger's parameterization technique:
\be
    \frac{1}{X^\Delta}=\frac{1}{\Gamma(\Delta)}\int_0^\infty d\zeta\,\zeta^{\Delta-1}\,e^{-X\zeta}\,.
\ee
In the fourth equality, the $\vec x'_\perp$ integration can be conducted by Gaussian integration and the $\tau'$ integration turns out to be $2\pi I_{\omega_E}(a\zeta)$ where $I_{\omega_E}(a\zeta)$ is a modified Bessel function of the first kind. In the fifth equality, the $\rho'$ integration gives $2 K_{ik}(b\zeta)\, K_{ik}(|\vec p_\perp|\rho)$. Finally, the last equality comes from the relation 
\be
    \int_0^\infty d\zeta\,\zeta^{\Delta-\frac{d}{2}}\,I_{\omega_E}(a\zeta)K_{ik}(b\zeta)=\frac{a^{\omega_E}|\Gamma(\alpha)|^2\;_2F_1\left(\alpha,\alpha^*;\alpha+\alpha^*+\frac{d}{2}-\Delta;\frac{a^2}{b^2}\right)}{b^{-\frac{d-2}2+\Delta+\omega_E}\,2^{\frac{d}{2}-\Delta+1}\Gamma(\alpha+\alpha^*+\frac{d}{2}-\Delta)}\,,
\ee
with the condition $\mathrm{Re}(\alpha)>0$\footnote{Note that we should be careful when $\mathrm{Re}(\alpha)\leq 0$. We leave this case as future work.} and $b>a$, which gives the result
\begin{align}
	&\mathcal{G}_{(a,b)}^\Delta(\omega_E,k)\nonumber\\
	&=\frac{\pi^\frac{d}{2}}{\Gamma(\Delta)}\frac{a^{\omega_E}}{b^{\Delta+\omega_E}}\frac{|\Gamma(\alpha)|^2}{\Gamma(\alpha+\alpha^*+\frac{d}{2}-\Delta)}\;_2F_1\left(\alpha,\alpha^*;\alpha+\alpha^*+\frac{d}{2}-\Delta\,;\frac{a^2}{b^2}\right)\,.\label{fttf}
\end{align}
In \eqref{fttf11}, we replaced $a\rightarrow z,b\rightarrow1$ so that only $z$ goes to $1^-$ limit.

\section{Useful expressions for the Hypergeometric function}\label{regularize}

To analyze the pole-skipping structure it is useful to express the Hypergeoemtric function in terms of gamma and digamma functions. In this appendix, we summarize various expressions for the Hypergeometric function that we have used in the main text. 

For the hypergeometric function ${}_2F_1(a,b;a+b+\mathcal{N};z)$ around $z=1$ \cite{NIST:DLMF} we consider four cases depending on whether $\mathcal{N}$ is a non-integer, a positive integer, zero or a negative integer. For all cases, we display the $z\rightarrow1^-$ limit (denoted by $\stackrel{z\rightarrow 1^-}{\simeq}$) in the last line where we only extract the finite terms, of order $(1-z)^0$.\footnote{In the cases where $\mathcal{N}$ is zero or a negative integer, we also discarded regular terms which are not relevant to the pole-skipping structure.} The Pochhammer symbol is defined as $(x)_j\equiv\frac{\Gamma(x+j)}{\Gamma(x)}=(x+j-1)\cdots(x+1)\cdot(x)$ for integer $j$.

\subsubsection*{If $\mathcal{N}$ is a non-integer ($\mathcal{N} \to n$): }
\begin{align}
&{}_2F_1(a,b;a+b+n;z)\nonumber\\
&= \frac{\Gamma(a+b+n)\Gamma(n)}{\Gamma(a+n) \Gamma(b+n)}\sum_{k=0}^\infty\frac{(a)_k(b)_k}{(1-n)_k k!}(1-z)^k\nonumber\\
&\quad+(1-z)^n\frac{\Gamma(a+b+n)\Gamma(-n)}{\Gamma(a) \Gamma(b)}\sum_{k=0}^\infty\frac{(a+n)_k(b+n)_k}{(1+n)_k k!}(1-z)^k\label{hypNon} \\
&\stackrel{z\rightarrow 1^-}{\simeq} \frac{\Gamma(a+b+n)\Gamma(n)}{\Gamma(a+n) \Gamma(b+n)}\,.\label{hypNon1}
\end{align}

\subsubsection*{If $\mathcal{N}$ is a positive integer ($\mathcal{N} \to N$):}
\begin{align}
\label{posIhyp}
&{}_2F_{1}(a,b;a+b+{N};z)\nonumber\\
&=\frac{\Gamma(a +b +N)}{\Gamma(a +N)\Gamma(b +N)} \sum_{k=0}^{N-1}\frac{(a )_k (b )_k (N-k-1)!}{k!}(z-1)^k\nonumber\\
&\quad-(z-1)^N \frac{\Gamma(a +b +N)}{\Gamma(a ) \Gamma(b )}\sum_{k=0}^{\infty} \frac{(a +N)_k(b +N)_k}{k!(k+N)!} (1-z)^k \times \nonumber \\
&\quad\left[\log(1-z)\right.-\psi(k+1)-
\left. \psi(k+N+1)+\psi(a +k+N)+\psi(b +k+N) \right]\\
&\stackrel{z\rightarrow 1^-}{\simeq}\frac{\Gamma(a +b +N)\Gamma(N)}{\Gamma(a+N ) \Gamma(b+N)}\,.\label{hypPI1}
\end{align}

\subsubsection*{If $\mathcal{N} = 0$:}
\begin{align} 
&{}_2F_{1}(a,b;a+b;z)\nonumber\\
&=- \frac{\Gamma(a +b )}{\Gamma(a ) \Gamma(b )}\sum_{k=0}^{\infty} \frac{(a )_k(b )_k}{(k!)^2} (1-z)^k\times\nonumber\\
&\quad\left[\log(1-z)-2\psi(k+1)+\psi(a +k)+\psi(b +k) \right]\label{hypZ}\\
&\stackrel{z\rightarrow 1^-}{\simeq}-\frac{\Gamma(a +b )}{\Gamma(a ) \Gamma(b)}\left[ \psi(a )+\psi(b ) \right]\label{hypZ1}\,.
\end{align}

\subsubsection*{If $\mathcal{N}$ is a negagive integer ($\mathcal{N} \to - N$):}
\begin{align}
&{}_2F_1(a,b;a+b-N;z)\nonumber\\
&=(1-z)^{-N}\frac{\Gamma(a +b -N)}{\Gamma(a )\Gamma(b )} \sum_{k=0}^{N-1}\frac{(a-N )_k (b-N )_k (N-k-1)!}{k!}(z-1)^k\nonumber\\
&\quad-(-1)^N \frac{\Gamma(a +b -N)}{\Gamma(a-N ) \Gamma(b-N )}\sum_{k=0}^{\infty} \frac{(a )_k(b )_k}{k!(k+N)!} (1-z)^k \times \nonumber \\
&\quad\left[\log(1-z)\right.-\psi(k+1)-
\left. \psi(k+N+1)+\psi(a +k)+\psi(b +k) \right]\label{hypNI}\\
&\stackrel{z\rightarrow 1^-}{\simeq}-\frac{(-1)^N}{N!} \frac{\Gamma(a +b -N)}{\Gamma(a-N ) \Gamma(b-N )}\left[ \psi(a )+\psi(b ) \right]\,.\label{hypNI1}
\end{align}
To obtain \eqref{hypNI}, we may use the relation
\begin{equation}
{}_2F_1(a,b;a+b-N;z)=(1-z)^{-N}{}_2F_1(a-N,b-N;a+b-N;z) \,,
\end{equation}
where the third argument has the form of $((a-N) + (b-N) + N)$ which allows us to use the expression for positive integers \eqref{posIhyp}.

In short, we have two replacement rules for the computation in Section \ref{section3}, which depend on the value of $\mathcal N$: i)  $\mathcal N$ is a non-integer or a positive integer or ii)  $\mathcal N$ is zero or a negative integer, 
\begin{align}
    \mathrm{i)\ } &{}_2F_1(a,b;a+b+\mathcal N;z) \stackrel{z\rightarrow 1^-}{\simeq} \frac{\Gamma(a+b+\mathcal N)\Gamma(\mathcal N)}{\Gamma(a+\mathcal N) \Gamma(b+\mathcal N)}    \label{rep1}\\
    \mathrm{ii)\ } &{}_2F_1(a,b;a+b+\mathcal N;z)\stackrel{z\rightarrow 1^-}{\simeq} \frac{\Gamma(a +b +\mathcal N)}{\Gamma(a+\mathcal N ) \Gamma(b+\mathcal N )}\left[ \psi(a )+\psi(b ) \right]    \label{rep2}\,,
\end{align}
where we removed an overall factor in the case ii) because it is irrelevant for our purposes. From these formulas, we also find a simple prescription for the case where $\mathcal{N}$ is zero or a negative integer: 1) we start with \eqref{rep1}, which is basically the formula for the non-integer $\mathcal{N}$ case; 2) if $\mathcal{N}$ is zero or a negative integer, $\Gamma(\mathcal{N})$ in \eqref{rep1} is divergent and should be replaced by the sum of two digamma functions whose arguments are inherited from $\Gamma$ functions. Symbolically, 
\begin{equation}
    \Gamma(\mathcal{N}) \to \psi(a) + \psi(b) \,. \label{presc1}
\end{equation}

\section{Hypergeometric functions: holographic perspective} \label{holoregul}

In the holographic computations, the bulk fields propagating on a Rindler-AdS$_{d+1}$ geometry have the following form
\be
\label{bulkFP}
(1-z)^{\frac{\Delta_{+}-p}{2}} {}_2F_1(a,b;a+b+\mathcal{N};z)\,,
\ee
where $\Delta_{+} = d/2 + \sqrt{(d-2p)^2+4m^2}/2$, with $p=0$ for scalar fields and $p=1$ for vector fields. Note that $\mathcal{N}$ is defined as
\begin{equation}
 \mathcal{N}:=d/2-\Delta_{+} \le 0 \,.
\end{equation}
  The retarded Green's function is obtained from the near-boundary ($z \approx 1$) behavior of the bulk fields, which takes the general form
\be
\label{genboundarybehav}
(1-z)^{\frac{\Delta_{+}-p}{2}}  \Big[A(1-z)^\mathcal{N} + \Big( B+C \log(1-z) \Big) \Big]\,,
\ee
where the value of the coefficients $A$, $B$ and $C$ depend on whether $\mathcal{N}$ takes integer values or not. In particular, $C = 0$ if $\mathcal{N}$ is a non-integer. We first consider the standard quantization, $\Delta = \Delta_+$ ($\Delta \ge d/2$), in three cases: negative non-integer $\mathcal{N}$, integer $\mathcal{N}$, and zero $\mathcal{N}$. 

\paragraph{Negative non-integer $\mathcal{N}$ case:}
The near boundary behavior ($z \to 1$) of the hypergeometric function in \eqref{bulkFP} is given by \eqref{hypNon}, from which, we may read off $A$ and $B$ in \eqref{genboundarybehav}:
\begin{equation}
\label{nIcoef}
A = \frac{\Gamma (-\mathcal{N}) \Gamma (a+b+\mathcal{N})}{\Gamma (a) \Gamma (b)}, \quad B = \frac{\Gamma (\mathcal{N}) \Gamma (a+b+\mathcal{N})}{\Gamma (a+\mathcal{N}) \Gamma (b+\mathcal{N})} \,.
\end{equation}
Thus, the Green's function reads
\be
\label{nIstand}
G^R(\omega,L)\propto \frac{B}{A} \propto  \frac{\Gamma(a)\Gamma(b)}{\Gamma(a+\mathcal{N})\Gamma(b+\mathcal{N})}\frac{\Gamma(\mathcal{N})}{\Gamma(-\mathcal{N})}
\,\,\,\,\text{for}\,\,\, \Delta = d/2 + n \ (n >0 \, \& \, n \notin \mathbb{N})\,, 
\ee
where $\mathbb{N}$ denotes the set of natural numbers. 

\paragraph{Negative integer $\mathcal{N}$ case:} The near boundary behavior of the hypergeometric function is given by \eqref{hypNI}. In this case $A$ and $B$ can be read off as
\begin{equation}
\label{Icoef}
    A = \frac{(-\mathcal{N}-1)!\Gamma (a+b+\mathcal{N})}{\Gamma (a) \Gamma (b)}, \quad B = -\frac{(-1)^{-\mathcal{N}}}{(-\mathcal{N})!}\frac{\Gamma (a+b+\mathcal{N})}{\Gamma (a+\mathcal{N}) \Gamma (b+\mathcal{N})}\big(\psi (a)+\psi (b)\big),
\end{equation}
which gives the  Green's function
\be
\label{Istand}
G^R(\omega,L) \propto \frac{B}{A} \propto \frac{\Gamma(a)\Gamma(b)}{\Gamma(a+\mathcal{N})\Gamma(b+\mathcal{N})}\,\big[\psi(a)+\psi(b) \big] \,\,\,\,\text{for}\,\,\, \Delta = d/2 + N \ (N \in \mathbb{N})\,. 
\ee

\paragraph{$\mathcal{N} = 0$ case:} 
By using \eqref{hypZ}, $A$ and $C$ can be read off as
\begin{equation} \label{xxx999}
    C = -\frac{\Gamma(a+b)}{\Gamma(a)\Gamma(b)}\,, \quad B = -\frac{\Gamma(a+b)}{\Gamma(a)\Gamma(b)}\big(\psi(a) + \psi(b)\big) \,, 
\end{equation}
and
\be
\label{zerocase}
G^R(\omega,L) \propto \frac{B}{C} \propto\psi(a)+\psi(b) \,\,\,\,\text{for}\,\,\, \Delta = d/2 \,.
\ee
Note that the simple relation    
\be
\frac{\Gamma(\mathcal{N})}{\Gamma(-\mathcal{N})} \to \psi(a)+\psi(b) \,,
\ee
can be used as a prescription to obtain \eqref{Istand} and \eqref{zerocase} from a formula such as \eqref{nIstand}. This relation corresponds to \eqref{presc1}.

Next, for $\Delta  < d/2$ cases, we need to consider the alternative quantization.  The mathematical results \eqref{nIcoef}, \eqref{Icoef} and \eqref{xxx999} with $\Delta_+$ are still valid no matter which quantization we consider. However, for the alternative quantization, we have to identify the conformal dimension $\Delta$ with $\Delta_- = d-\Delta_+$, not $\Delta_+$. 

For example, let us consider the scalar field \eqref{scalarsol}:
\begin{equation}
    a(\Delta_{+}) = \,\frac{\Delta_{+}+2-d-L-i \omega}{2}, \quad b(\Delta_{+}) = \frac{\Delta_{+}+L-i \omega}{2}, \quad \mathcal{N}(\Delta_{+}) = \frac{d}{2}-\Delta_{+}\,,
\end{equation}
where we replace $\{a,b,\mathcal{N}\}\rightarrow \{a(\Delta_{+}),b(\Delta_{+}),\mathcal{N}(\Delta_{+})\}$, to emphasize their $\Delta_{+}$ dependence. In this example, we have the following relations
\begin{equation}
\label{rels}
\begin{split}
    &\mathcal{N}(\Delta_{+}) = -\mathcal{N}(\Delta_{-})\,,  \\
    &a(\Delta_{+}) = a(\Delta_{-})+\mathcal{N}(\Delta_{-})\,, \\
    &b(\Delta_{+}) = b(\Delta_{-})+\mathcal{N}(\Delta_{-})\,.
    \end{split}
\end{equation}
By using \eqref{rels}, $A$ and $B$ in \eqref{nIcoef} can be written in terms of $\{a(\Delta_{-}),b(\Delta_{-}),n(\Delta_{-})\}$ as
\begin{equation}
\label{nIcoefalter}
A = \frac{\Gamma (\mathcal{N}) \Gamma (a+b+\mathcal{N})}{\Gamma (a+\mathcal{N}) \Gamma (b+\mathcal{N})}\,, \quad B = \frac{\Gamma (-\mathcal{N}) \Gamma (a+b+\mathcal{N})}{\Gamma (a) \Gamma (b)}\,,
\end{equation}
where we omit the $\Delta_{-}$ dependence. Hence, in the alternative quantization scheme, the retarded Green's function of scalar fields for non-integer $\mathcal{N}>0$ is
\begin{equation}
\label{nIalter}
    G^R(\omega,L)\propto \frac{A}{B} \propto \frac{\Gamma(\mathcal{N})}{\Gamma(-\mathcal{N})} \frac{\Gamma(a)\Gamma(b)}{\Gamma(a+\mathcal{N})\Gamma(b+\mathcal{N})}\,\,\,\,\text{for}\,\,\, \Delta = d/2 - n \ (n >0 \, \& \, n \notin \mathbb{N})\,, 
\end{equation}
which is the same as \eqref{nIstand}.


\section{Hyperspherical harmonics}\label{ap-hyper}

The hyperspherical harmonics $Y_{L\,K}^{(d-1)}$ are eigenfunctions of the Laplace-Beltrami operator on $S^{d-1}$, i.e.,
\be \label{eq-laplace}
\square_{S^{d-1}} Y_{L\,K}^{(d-1)}(\theta_i)=-L (L+d-2)Y_{L\,K}^{(d-1)}(\theta_i)\,,
\ee
where $\theta_i \in S^{d-1}$ and $L=0,1,2,...$ is the generalized angular momentum quantum number. $K$ stands for a set of $d-2$ quantum numbers identifying degenerate harmonics for each $L$\footnote{For more details about hyperspherical harmonics, see for instance \cite{hyper}.}. Below, we will also use the notation $L=\mu_1$ and $K=(\mu_2, \mu_3, ...,\mu_{d-1})$.

To find the eigenfucntions of the Laplace-Beltrami operator on $\mathbb H^{d-1}$, we first write the metric on $S^{d-1}$ as $ds^2=d\theta^2 +\sin^2 \, \theta d\Omega_{d-2}^2$. Under the analytic continuation $\theta = i \chi$, the operator $ \square_{S^{d-1}}$ becomes $-\square_{\mathbb H^{d-1}}$, and (\ref{eq-laplace}) can be written as 
\be \label{c120}
\square_{\mathbb H^{d-1}} Y_{L\,K}^{(d-1)}( i \chi,\phi_i)=L (L+d-2)Y_{L\,K}^{(d-1)}(i \chi,\phi_i)\,,
\ee
where $\phi_i \in S^{d-2}$. The angular momentum quantum number $L$ can be extended to take non-integer values by writing the hyperspherical harmonics $Y_{L\,K}^{(d-1)}$ in terms of hypergeometric functions. To do that, one first writes the hyperspherical harmonics in terms of Gegenbauer functions $C_{\nu}^{(\alpha)}(z)$ \cite{hyper}
\be
Y_{\mu_1,\mu_2,...} \sim e^{i m \phi} \prod_{j=1}^{d-2}C_{\mu_j-\mu_{j+1}}^{\left( \frac{d-j-2}{2}+\mu_{j+1}\right)}(\cos \theta_j)(\sin \theta_j)^{\mu_j+1}\,,
\ee
and then one writes the Gegenbauer functions in terms of hypergeometric functions
\be
C_{\nu}^{(\alpha)}(z)=\frac{2^{1-2\alpha}\sqrt{\pi}\Gamma(\nu+2\alpha)}{\nu! \Gamma(\alpha)} {}_2F_1\left(-n,2\alpha+n;\alpha+\frac{1}{2};\frac{1-z}{2} \right)\,.
\ee
The final result is then an analytical function of $\nu, \alpha$ and $z$, which allows us to define $Y_{\mu_1,\mu_2,...}$ for arbitrary values of the parameters $\mu_j$.


\bibliography{HyunSikRefs}

\providecommand{\href}[2]{#2}\begingroup\raggedright\begin{thebibliography}{10}

\bibitem{larkin1969quasiclassical}
A.~Larkin and Y.~N. Ovchinnikov, \emph{Quasiclassical method in the theory of
  superconductivity}, {\emph{Sov Phys JETP} {\bf 28} (1969) 1200--1205}.

\bibitem{Kitaev-2014}
A.~Kitaev, \emph{{A simple model of quantum holography}}, {\emph{{}} (2015)
  \url{http://online.kitp.ucsb.edu/online/entangled15/kitaev/},
  \url{http://online.kitp.ucsb.edu/online/entangled15/kitaev2/}, Talks at KITP,
  April 7, 2015 and May 27, (2015)}.

\bibitem{Maldacena:2015waa}
J.~Maldacena, S.~H. Shenker and D.~Stanford, \emph{{A bound on chaos}},
  \href{http://dx.doi.org/10.1007/JHEP08(2016)106}{\emph{JHEP} {\bf 08} (2016)
  106}, [\href{http://arxiv.org/abs/1503.01409}{{\tt 1503.01409}}].

\bibitem{Swingle:2016jdj}
B.~Swingle and D.~Chowdhury, \emph{{Slow scrambling in disordered quantum
  systems}}, \href{http://dx.doi.org/10.1103/PhysRevB.95.060201}{\emph{Phys.
  Rev.} {\bf B95} (2017) 060201}, [\href{http://arxiv.org/abs/1608.03280}{{\tt
  1608.03280}}].

\bibitem{Aleiner:2016eni}
I.~L. Aleiner, L.~Faoro and L.~B. Ioffe, \emph{{Microscopic model of quantum
  butterfly effect: out-of-time-order correlators and traveling combustion
  waves}}, \href{http://dx.doi.org/10.1016/j.aop.2016.09.006}{\emph{Annals
  Phys.} {\bf 375} (2016) 378--406},
  [\href{http://arxiv.org/abs/1609.01251}{{\tt 1609.01251}}].

\bibitem{Xu:2018xfz}
S.~Xu and B.~Swingle, \emph{{Accessing scrambling using matrix product
  operators}}, \href{http://dx.doi.org/10.1038/s41567-019-0712-4}{\emph{Nature
  Phys.} {\bf 16} (2019) 199--204},
  [\href{http://arxiv.org/abs/1802.00801}{{\tt 1802.00801}}].

\bibitem{Khemani:2018sdn}
V.~Khemani, D.~A. Huse and A.~Nahum, \emph{{Velocity-dependent Lyapunov
  exponents in many-body quantum, semiclassical, and classical chaos}},
  \href{http://dx.doi.org/10.1103/PhysRevB.98.144304}{\emph{Phys. Rev.} {\bf
  B98} (2018) 144304}, [\href{http://arxiv.org/abs/1803.05902}{{\tt
  1803.05902}}].

\bibitem{Roberts:2014ifa}
D.~A. Roberts and D.~Stanford, \emph{{Two-dimensional conformal field theory
  and the butterfly effect}},
  \href{http://dx.doi.org/10.1103/PhysRevLett.115.131603}{\emph{Phys. Rev.
  Lett.} {\bf 115} (2015) 131603}, [\href{http://arxiv.org/abs/1412.5123}{{\tt
  1412.5123}}].

\bibitem{Shenker:2014cwa}
S.~H. Shenker and D.~Stanford, \emph{{Stringy effects in scrambling}},
  \href{http://dx.doi.org/10.1007/JHEP05(2015)132}{\emph{JHEP} {\bf 05} (2015)
  132}, [\href{http://arxiv.org/abs/1412.6087}{{\tt 1412.6087}}].

\bibitem{Shenker:2013pqa}
S.~H. Shenker and D.~Stanford, \emph{{Black holes and the butterfly effect}},
  \href{http://dx.doi.org/10.1007/JHEP03(2014)067}{\emph{JHEP} {\bf 03} (2014)
  067}, [\href{http://arxiv.org/abs/1306.0622}{{\tt 1306.0622}}].

\bibitem{Shenker:2013yza}
S.~H. Shenker and D.~Stanford, \emph{{Multiple Shocks}},
  \href{http://dx.doi.org/10.1007/JHEP12(2014)046}{\emph{JHEP} {\bf 12} (2014)
  046}, [\href{http://arxiv.org/abs/1312.3296}{{\tt 1312.3296}}].

\bibitem{Roberts:2014isa}
D.~A. Roberts, D.~Stanford and L.~Susskind, \emph{{Localized shocks}},
  \href{http://dx.doi.org/10.1007/JHEP03(2015)051}{\emph{JHEP} {\bf 03} (2015)
  051}, [\href{http://arxiv.org/abs/1409.8180}{{\tt 1409.8180}}].

\bibitem{Polchinski:2016xgd}
J.~Polchinski and V.~Rosenhaus, \emph{{The Spectrum in the Sachdev-Ye-Kitaev
  Model}}, \href{http://dx.doi.org/10.1007/JHEP04(2016)001}{\emph{JHEP} {\bf
  04} (2016) 001}, [\href{http://arxiv.org/abs/1601.06768}{{\tt 1601.06768}}].

\bibitem{Perlmutter:2016pkf}
E.~Perlmutter, \emph{{Bounding the Space of Holographic CFTs with Chaos}},
  \href{http://dx.doi.org/10.1007/JHEP10(2016)069}{\emph{JHEP} {\bf 10} (2016)
  069}, [\href{http://arxiv.org/abs/1602.08272}{{\tt 1602.08272}}].

\bibitem{Maldacena:2016hyu}
J.~Maldacena and D.~Stanford, \emph{{Remarks on the Sachdev-Ye-Kitaev model}},
  \href{http://dx.doi.org/10.1103/PhysRevD.94.106002}{\emph{Phys. Rev.} {\bf
  D94} (2016) 106002}, [\href{http://arxiv.org/abs/1604.07818}{{\tt
  1604.07818}}].

\bibitem{Jensen:2016pah}
K.~Jensen, \emph{{Chaos in AdS$_2$ Holography}},
  \href{http://dx.doi.org/10.1103/PhysRevLett.117.111601}{\emph{Phys. Rev.
  Lett.} {\bf 117} (2016) 111601}, [\href{http://arxiv.org/abs/1605.06098}{{\tt
  1605.06098}}].

\bibitem{Maldacena:2016upp}
J.~Maldacena, D.~Stanford and Z.~Yang, \emph{{Conformal symmetry and its
  breaking in two dimensional Nearly Anti-de-Sitter space}},
  \href{http://dx.doi.org/10.1093/ptep/ptw124}{\emph{PTEP} {\bf 2016} (2016)
  12C104}, [\href{http://arxiv.org/abs/1606.01857}{{\tt 1606.01857}}].

\bibitem{Davison:2016ngz}
R.~A. Davison, W.~Fu, A.~Georges, Y.~Gu, K.~Jensen and S.~Sachdev,
  \emph{{Thermoelectric transport in disordered metals without quasiparticles:
  The Sachdev-Ye-Kitaev models and holography}},
  \href{http://dx.doi.org/10.1103/PhysRevB.95.155131}{\emph{Phys. Rev. B} {\bf
  95} (2017) 155131}, [\href{http://arxiv.org/abs/1612.00849}{{\tt
  1612.00849}}].

\bibitem{Jahnke:2018off}
V.~Jahnke, \emph{{Recent developments in the holographic description of quantum
  chaos}}, \href{http://dx.doi.org/10.1155/2019/9632708}{\emph{Adv. High Energy
  Phys.} {\bf 2019} (2019) 9632708},
  [\href{http://arxiv.org/abs/1811.06949}{{\tt 1811.06949}}].

\bibitem{Jahnke:2019gxr}
V.~Jahnke, K.-Y. Kim and J.~Yoon, \emph{{On the Chaos Bound in Rotating Black
  Holes}}, \href{http://dx.doi.org/10.1007/JHEP05(2019)037}{\emph{JHEP} {\bf
  05} (2019) 037}, [\href{http://arxiv.org/abs/1903.09086}{{\tt 1903.09086}}].

\bibitem{Fischler:2018kwt}
W.~Fischler, V.~Jahnke and J.~F. Pedraza, \emph{{Chaos and entanglement
  spreading in a non-commutative gauge theory}},
  \href{http://dx.doi.org/10.1007/JHEP11(2018)072}{\emph{JHEP} {\bf 11} (2018)
  072}, [\href{http://arxiv.org/abs/1808.10050}{{\tt 1808.10050}}].

\bibitem{Avila:2018sqf}
D.~Avila, V.~Jahnke and L.~Pati\~{n}o, \emph{{Chaos, Diffusivity, and Spreading
  of Entanglement in Magnetic Branes, and the Strengthening of the Internal
  Interaction}}, \href{http://dx.doi.org/10.1007/JHEP09(2018)131}{\emph{JHEP}
  {\bf 09} (2018) 131}, [\href{http://arxiv.org/abs/1805.05351}{{\tt
  1805.05351}}].

\bibitem{Jahnke:2017iwi}
V.~Jahnke, \emph{{Delocalizing entanglement of anisotropic black branes}},
  \href{http://dx.doi.org/10.1007/JHEP01(2018)102}{\emph{JHEP} {\bf 01} (2018)
  102}, [\href{http://arxiv.org/abs/1708.07243}{{\tt 1708.07243}}].

\bibitem{Aalsma:2020aib}
L.~Aalsma and G.~Shiu, \emph{{Chaos and complementarity in de Sitter space}},
  \href{http://dx.doi.org/10.1007/JHEP05(2020)152}{\emph{JHEP} {\bf 05} (2020)
  152}, [\href{http://arxiv.org/abs/2002.01326}{{\tt 2002.01326}}].

\bibitem{Geng:2020kxh}
H.~Geng, \emph{{Non-local Entanglement and Fast Scrambling in De-Sitter
  Holography}},  \href{http://arxiv.org/abs/2005.00021}{{\tt 2005.00021}}.

\bibitem{Grozdanov:2017ajz}
S.~Grozdanov, K.~Schalm and V.~Scopelliti, \emph{{Black hole scrambling from
  hydrodynamics}},
  \href{http://dx.doi.org/10.1103/PhysRevLett.120.231601}{\emph{Phys. Rev.
  Lett.} {\bf 120} (2018) 231601}, [\href{http://arxiv.org/abs/1710.00921}{{\tt
  1710.00921}}].

\bibitem{Blake:2017ris}
M.~Blake, H.~Lee and H.~Liu, \emph{{A quantum hydrodynamical description for
  scrambling and many-body chaos}},
  \href{http://dx.doi.org/10.1007/JHEP10(2018)127}{\emph{JHEP} {\bf 10} (2018)
  127}, [\href{http://arxiv.org/abs/1801.00010}{{\tt 1801.00010}}].

\bibitem{Blake:2018leo}
M.~Blake, R.~A. Davison, S.~Grozdanov and H.~Liu, \emph{{Many-body chaos and
  energy dynamics in holography}},
  \href{http://dx.doi.org/10.1007/JHEP10(2018)035}{\emph{JHEP} {\bf 10} (2018)
  035}, [\href{http://arxiv.org/abs/1809.01169}{{\tt 1809.01169}}].

\bibitem{Grozdanov:2019uhi}
S.~Grozdanov, P.~K. Kovtun, A.~O. Starinets and P.~Tadi\'{c}, \emph{{The
  complex life of hydrodynamic modes}},
  \href{http://dx.doi.org/10.1007/JHEP11(2019)097}{\emph{JHEP} {\bf 11} (2019)
  097}, [\href{http://arxiv.org/abs/1904.12862}{{\tt 1904.12862}}].

\bibitem{Blake:2019otz}
M.~Blake, R.~A. Davison and D.~Vegh, \emph{{Horizon constraints on holographic
  Green's functions}},
  \href{http://dx.doi.org/10.1007/JHEP01(2020)077}{\emph{JHEP} {\bf 01} (2020)
  077}, [\href{http://arxiv.org/abs/1904.12883}{{\tt 1904.12883}}].

\bibitem{Natsuume:2019xcy}
M.~Natsuume and T.~Okamura, \emph{{Nonuniqueness of Green's functions at
  special points}},
  \href{http://dx.doi.org/10.1007/JHEP12(2019)139}{\emph{JHEP} {\bf 12} (2019)
  139}, [\href{http://arxiv.org/abs/1905.12015}{{\tt 1905.12015}}].

\bibitem{Natsuume:2019vcv}
M.~Natsuume and T.~Okamura, \emph{{Pole-skipping with finite-coupling
  corrections}},
  \href{http://dx.doi.org/10.1103/PhysRevD.100.126012}{\emph{Phys. Rev.} {\bf
  D100} (2019) 126012}, [\href{http://arxiv.org/abs/1909.09168}{{\tt
  1909.09168}}].

\bibitem{Wu:2019esr}
X.~Wu, \emph{{Higher curvature corrections to pole-skipping}},
  \href{http://dx.doi.org/10.1007/JHEP12(2019)140}{\emph{JHEP} {\bf 12} (2019)
  140}, [\href{http://arxiv.org/abs/1909.10223}{{\tt 1909.10223}}].

\bibitem{Abbasi:2019rhy}
N.~Abbasi and J.~Tabatabaei, \emph{{Quantum chaos, pole-skipping and
  hydrodynamics in a holographic system with chiral anomaly}},
  \href{http://dx.doi.org/10.1007/JHEP03(2020)050}{\emph{JHEP} {\bf 03} (2020)
  050}, [\href{http://arxiv.org/abs/1910.13696}{{\tt 1910.13696}}].

\bibitem{Grozdanov:2018kkt}
S.~Grozdanov, \emph{{On the connection between hydrodynamics and quantum chaos
  in holographic theories with stringy corrections}},
  \href{http://dx.doi.org/10.1007/JHEP01(2019)048}{\emph{JHEP} {\bf 01} (2019)
  048}, [\href{http://arxiv.org/abs/1811.09641}{{\tt 1811.09641}}].

\bibitem{Li:2019bgc}
W.~Li, S.~Lin and J.~Mei, \emph{{Thermal diffusion and quantum chaos in neutral
  magnetized plasma}},
  \href{http://dx.doi.org/10.1103/PhysRevD.100.046012}{\emph{Phys. Rev. D} {\bf
  100} (2019) 046012}, [\href{http://arxiv.org/abs/1905.07684}{{\tt
  1905.07684}}].

\bibitem{Natsuume:2019sfp}
M.~Natsuume and T.~Okamura, \emph{{Holographic chaos, pole-skipping, and
  regularity}}, \href{http://dx.doi.org/10.1093/ptep/ptz155}{\emph{PTEP} {\bf
  2020} (2020) 013B07}, [\href{http://arxiv.org/abs/1905.12014}{{\tt
  1905.12014}}].

\bibitem{Ahn:2019rnq}
Y.~Ahn, V.~Jahnke, H.-S. Jeong and K.-Y. Kim, \emph{{Scrambling in Hyperbolic
  Black Holes: shock waves and pole-skipping}},
  \href{http://dx.doi.org/10.1007/JHEP10(2019)257}{\emph{JHEP} {\bf 10} (2019)
  257}, [\href{http://arxiv.org/abs/1907.08030}{{\tt 1907.08030}}].

\bibitem{Ceplak:2019ymw}
N.~Ceplak, K.~Ramdial and D.~Vegh, \emph{{Fermionic pole-skipping in
  holography}},  \href{http://arxiv.org/abs/1910.02975}{{\tt 1910.02975}}.

\bibitem{Liu:2020yaf}
Y.~Liu and A.~Raju, \emph{{Quantum Chaos in Topologically Massive Gravity}},
  \href{http://arxiv.org/abs/2005.08508}{{\tt 2005.08508}}.

\bibitem{Haehl:2018izb}
F.~M. Haehl and M.~Rozali, \emph{{Effective Field Theory for Chaotic CFTs}},
  \href{http://dx.doi.org/10.1007/JHEP10(2018)118}{\emph{JHEP} {\bf 10} (2018)
  118}, [\href{http://arxiv.org/abs/1808.02898}{{\tt 1808.02898}}].

\bibitem{Das:2019tga}
S.~Das, B.~Ezhuthachan and A.~Kundu, \emph{{Real time dynamics from low point
  correlators in 2d BCFT}},
  \href{http://dx.doi.org/10.1007/JHEP12(2019)141}{\emph{JHEP} {\bf 12} (2019)
  141}, [\href{http://arxiv.org/abs/1907.08763}{{\tt 1907.08763}}].

\bibitem{Haehl:2019eae}
F.~M. Haehl, W.~Reeves and M.~Rozali, \emph{{Reparametrization modes, shadow
  operators, and quantum chaos in higher-dimensional CFTs}},
  \href{http://dx.doi.org/10.1007/JHEP11(2019)102}{\emph{JHEP} {\bf 11} (2019)
  102}, [\href{http://arxiv.org/abs/1909.05847}{{\tt 1909.05847}}].

\bibitem{Camanho:2014apa}
X.~O. Camanho, J.~D. Edelstein, J.~Maldacena and A.~Zhiboedov, \emph{{Causality
  Constraints on Corrections to the Graviton Three-Point Coupling}},
  \href{http://dx.doi.org/10.1007/JHEP02(2016)020}{\emph{JHEP} {\bf 02} (2016)
  020}, [\href{http://arxiv.org/abs/1407.5597}{{\tt 1407.5597}}].

\bibitem{Heemskerk:2009pn}
I.~Heemskerk, J.~Penedones, J.~Polchinski and J.~Sully, \emph{{Holography from
  Conformal Field Theory}},
  \href{http://dx.doi.org/10.1088/1126-6708/2009/10/079}{\emph{JHEP} {\bf 10}
  (2009) 079}, [\href{http://arxiv.org/abs/0907.0151}{{\tt 0907.0151}}].

\bibitem{Cornalba:2006xm}
L.~Cornalba, M.~S. Costa, J.~Penedones and R.~Schiappa, \emph{{Eikonal
  Approximation in AdS/CFT: Conformal Partial Waves and Finite N Four-Point
  Functions}},
  \href{http://dx.doi.org/10.1016/j.nuclphysb.2007.01.007}{\emph{Nucl. Phys.}
  {\bf B767} (2007) 327--351}, [\href{http://arxiv.org/abs/hep-th/0611123}{{\tt
  hep-th/0611123}}].

\bibitem{Casini:2011kv}
H.~Casini, M.~Huerta and R.~C. Myers, \emph{{Towards a derivation of
  holographic entanglement entropy}},
  \href{http://dx.doi.org/10.1007/JHEP05(2011)036}{\emph{JHEP} {\bf 05} (2011)
  036}, [\href{http://arxiv.org/abs/1102.0440}{{\tt 1102.0440}}].

\bibitem{Simmons-Duffin:2016gjk}
D.~Simmons-Duffin, \emph{{The Conformal Bootstrap}},  in \emph{{Proceedings,
  Theoretical Advanced Study Institute in Elementary Particle Physics: New
  Frontiers in Fields and Strings (TASI 2015): Boulder, CO, USA, June 1-26,
  2015}}, pp.~1--74, 2017.
\newblock \href{http://arxiv.org/abs/1602.07982}{{\tt 1602.07982}}.
\newblock \href{http://dx.doi.org/10.1142/9789813149441_0001}{DOI}.

\bibitem{Dolan:2011dv}
F.~A. Dolan and H.~Osborn, \emph{{Conformal Partial Waves: Further Mathematical
  Results}},  \href{http://arxiv.org/abs/1108.6194}{{\tt 1108.6194}}.

\bibitem{NIST:DLMF}
``{\it NIST Digital Library of Mathematical Functions}.''
  http://dlmf.nist.gov/, Release 1.0.24 of 2019-09-15.

\bibitem{Ferrara:1972uq}
S.~Ferrara, A.~F. Grillo, G.~Parisi and R.~Gatto, \emph{{The shadow operator
  formalism for conformal algebra. Vacuum expectation values and operator
  products}}, \href{http://dx.doi.org/10.1007/BF02907130}{\emph{Lett. Nuovo
  Cim.} {\bf 4S2} (1972) 115--120}.

\bibitem{SimmonsDuffin:2012uy}
D.~Simmons-Duffin, \emph{{Projectors, Shadows, and Conformal Blocks}},
  \href{http://dx.doi.org/10.1007/JHEP04(2014)146}{\emph{JHEP} {\bf 04} (2014)
  146}, [\href{http://arxiv.org/abs/1204.3894}{{\tt 1204.3894}}].

\bibitem{Ohya:2016gto}
S.~Ohya, \emph{{Intertwining operator in thermal CFT$_d$}},
  \href{http://dx.doi.org/10.1142/S0217751X17500063}{\emph{Int. J. Mod. Phys.}
  {\bf A32} (2017) 1750006}, [\href{http://arxiv.org/abs/1611.00763}{{\tt
  1611.00763}}].

\bibitem{Costa:2011mg}
M.~S. Costa, J.~Penedones, D.~Poland and S.~Rychkov, \emph{{Spinning Conformal
  Correlators}}, \href{http://dx.doi.org/10.1007/JHEP11(2011)071}{\emph{JHEP}
  {\bf 11} (2011) 071}, [\href{http://arxiv.org/abs/1107.3554}{{\tt
  1107.3554}}].

\bibitem{pspcomments}
Y.~Ahn, V.~Jahnke, H.-S. Jeong, K.-Y. Kim, K.-S. Lee and M.~Nishida,
  \emph{{Classifying pole-skipping points}},
  \href{http://arxiv.org/abs/2010.16166}{{\tt 2010.16166}}.

\bibitem{Ueda:2018xvl}
K.~Ueda and A.~Ishibashi, \emph{{Massive vector field perturbations on extremal
  and near-extremal static black holes}},
  \href{http://dx.doi.org/10.1103/PhysRevD.97.124050}{\emph{Phys. Rev.} {\bf
  D97} (2018) 124050}, [\href{http://arxiv.org/abs/1805.02479}{{\tt
  1805.02479}}].

\bibitem{Kodama:2003kk}
H.~Kodama and A.~Ishibashi, \emph{{Master equations for perturbations of
  generalized static black holes with charge in higher dimensions}},
  \href{http://dx.doi.org/10.1143/PTP.111.29}{\emph{Prog. Theor. Phys.} {\bf
  111} (2004) 29--73}, [\href{http://arxiv.org/abs/hep-th/0308128}{{\tt
  hep-th/0308128}}].

\bibitem{Kodama:2003jz}
H.~Kodama and A.~Ishibashi, \emph{{A Master equation for gravitational
  perturbations of maximally symmetric black holes in higher dimensions}},
  \href{http://dx.doi.org/10.1143/PTP.110.701}{\emph{Prog. Theor. Phys.} {\bf
  110} (2003) 701--722}, [\href{http://arxiv.org/abs/hep-th/0305147}{{\tt
  hep-th/0305147}}].

\bibitem{Kodama:2000fa}
H.~Kodama, A.~Ishibashi and O.~Seto, \emph{{Brane world cosmology: Gauge
  invariant formalism for perturbation}},
  \href{http://dx.doi.org/10.1103/PhysRevD.62.064022}{\emph{Phys. Rev.} {\bf
  D62} (2000) 064022}, [\href{http://arxiv.org/abs/hep-th/0004160}{{\tt
  hep-th/0004160}}].

\bibitem{silverman1972special}
R.~A. Silverman et~al., \emph{Special functions and their applications}.
\newblock Courier Corporation, 1972.

\bibitem{hyper}
J.~A. J.E.~Avery, \emph{Hyperspherical Harmonics and Their Physical
  Applications}.
\newblock World Scientific, 2018.

\end{thebibliography}\endgroup
\bibliographystyle{JHEP}

\end{document}